\documentclass[footinbib,aip,apl,reprint]{revtex4-1}

\usepackage{graphicx}
\usepackage{amsmath}
\usepackage{mathtools}
\usepackage[absolute]{textpos}
\usepackage{comment}
\usepackage{bbm}

\begin{document}

\title{Reconfigurable generation and measurement of mutually unbiased bases for time-bin qudits}

\author{Joseph M. Lukens}
\email{lukensjm@ornl.gov}
\affiliation{Quantum Information Science Group, Computational Sciences and Engineering Division, Oak Ridge National Laboratory, Oak Ridge, Tennessee 37831, USA}

\author{Nurul T. Islam}
\affiliation{Department of Physics and Fitzpatrick Institute for Photonics, Duke University, Durham, North Carolina 27708, USA}

\author{Charles Ci Wen Lim}
\affiliation{Department of Electrical and Computer Engineering and Centre for Quantum Technologies, National University of Singapore, Singapore 117583, Singapore}

\author{Daniel J. Gauthier}
\affiliation{Department of Physics, The Ohio State University, Columbus, Ohio 43210, USA}

\date{\today}

\begin{abstract}
We propose a new method for implementing mutually unbiased generation and measurement of time-bin qudits using a cascade of electro-optic phase modulator -- coded fiber Bragg grating pairs. Our approach requires only a single spatial mode and can switch rapidly between basis choices. We obtain explicit solutions for dimensions $d=2,3,4$ that realize all $d+1$ possible mutually unbiased bases and analyze the performance of our approach in quantum key distribution. Given its practicality and compatibility with current technology, our approach provides a promising springboard for scalable processing of high-dimensional time-bin states.
\end{abstract}

\maketitle

\begin{textblock}{13.45}(1.35,15)
\noindent \fontsize{7}{7}\selectfont This manuscript has been authored by UT-Battelle, LLC under Contract No. DE-AC05-00OR22725 with the U.S. Department of Energy. The United States Government retains and the publisher, by accepting the article for publication, acknowledges that the United States Government retains a non-exclusive, paid-up, irrevocable, world-wide license to publish or reproduce the published form of this manuscript, or allow others to do so, for United States Government purposes. The Department of Energy will provide public access to these results of federally sponsored research in accordance with the DOE Public Access Plan. (http://energy.gov/downloads/doe-public-access-plan).
\end{textblock}

Mutually unbiased bases (MUBs) are of fundamental importance in quantum mechanics. A collection of bases is called mutually unbiased if any eigenstate from one basis overlaps equally with all states from the other bases. Since a measurement result in one basis provides no predictive information about the outcome of a subsequent measurement in one of its mutually unbiased partners,~\cite{Wootters1989, Durt2010} MUBs represent optimal choices for quantum state tomography in noisy environments~\cite{Wootters1989, Adamson2010} and guarantee security against eavesdropping in quantum key distribution (QKD).~\cite{Gisin2002, Scarani2009} Implementing MUBs experimentally can be challenging, especially in high-dimensional Hilbert spaces (dimension $d>2$). For example, in time-bin encoding, the best known method for measuring states mutually-unbiased with respect to single-time-bin wavepackets is to use nested delay interferometers (DIs).~\cite{Brougham2013a,Islam2017a,Islam2017b} This approach requires the use of $d$ detectors and is difficult to scale to high $d$ in a practical setting.  Furthermore, passive DIs produce satellite pulses that reduce the probability of successful measurement to $1/d$.~\cite{Hillerkuss2010, Hillerkuss2011}

\begin{figure}
\includegraphics[width=3.4in]{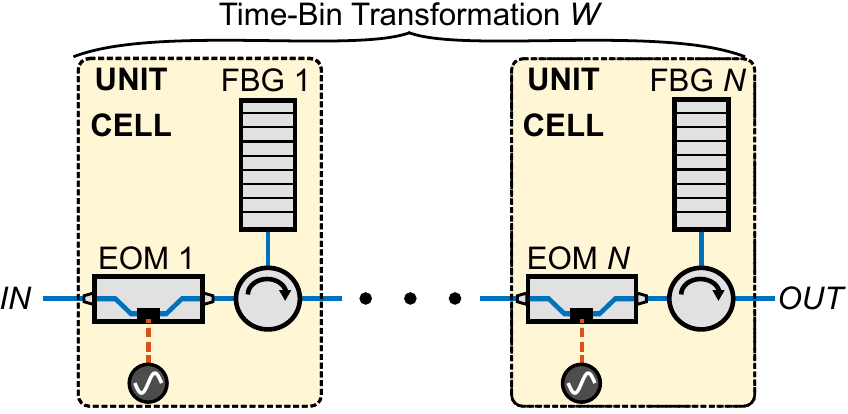}
\caption{Proposed system. The input state propagates through a series of $N$ unit cells, each comprising an EOM and coded FBG. Ideally the input-to-output path produces a one-to-one mapping between states from different bases.}
\label{fig1}
\end{figure}

Therefore, a critical objective is to develop an approach for synthesizing time-bin MUBs that preserves single-spatial-mode quantum information processing, is rapidly reconfigurable, and enables measurements with high success probability. Here we propose and analyze a novel configuration for generating and measuring quantum photonic states prepared in various time-bin MUBs using a cascade of electro-optic phase modulator (EOM) -- coded fiber Bragg grating (FBG) pairs (see Fig.~\ref{fig1}). We find explicit solutions of $(d+1)$-element MUB families for $d=2,3,4$ and simulate performance of a multibasis QKD system. Importantly, our approach can be implemented using current technology and offers the first truly single-mode paradigm for time-bin MUBs.

In our design, we impose the requirement that a physical configuration must transform between all MUBs available in a $d$-dimensional space. This criterion is motivated by high-dimensional QKD, as both increasing the dimension $d$ and employing $(d+1)$-basis protocols enhance robustness against noise.~\cite{Cerf2002, Sheridan2010} Moreover, we seek solutions that can efficiently detect each eigenstate in a MUB simultaneously, as opposed to projective measurements that only result in a binary outcome.~\cite{Lima2011, Giovannini2013} The key technology we explore for generating and measuring discrete time-bin states in multiple MUBs is the coded FBG, which has played an important role in the development of classical optical code division multiple access networks.~\cite{Yin2007} Such an FBG typically consists of a series of relatively low-reflectivity segments, each with a specified amplitude and phase shift. The (assumed fixed) length of each segment, or ``chip,'' sets the time-bin duration, so that the total reflected field for a single quantum wavepacket consists of a series of wavepackets with weight coefficients given by the chip sequence. By extension, if instead a train of wavepackets is input to the FBG, the FBG will cause interference between different time bins. Codes as long as 1023 chips have been implemented experimentally,~\cite{Dai2007} thus making FBGs an appealing choice for time-bin interference of quantum states in optical fiber. 

Unfortunately, coded gratings alone are insufficient for arbitrary mode transformations because they are linear, time-invariant filters, making it impossible to map unique interference patterns to each output bin. For example, a single FBG (or cascade thereof) can, in principle, produce one of the $d$ output transformations realized by $d-1$ nested DIs,~\cite{Chen2009} but it cannot replace all of them. This problem can be overcome by cascading the coded FBG with an additional time-dependent optical primitive: an electro-optic phase modulator (EOM), which enables user-defined phase shifts to any sequence of time bins.~\cite{Wooten2000} When integrated as a sequence of alternating FBGs and EOMs (Fig.~\ref{fig1}), it is possible to realize arbitrary temporal transformations where the number of components scales linearly with the dimensionality $d$---a conclusion reached by extending, via Fourier duality, recent results on frequency-bin quantum information processing.~\cite{Lukens2017} Nonetheless, it remains to be shown that a particular EOM/FBG sequence can switch between \emph{multiple} MUBs rapidly, a requirement which goes beyond the sufficiency of FBGs and EOMs to produce a \emph{given} transformation, and requires a different design procedure.

To formulate this problem, we assume a Hilbert space spanned by the single-photon states $|k\rangle = \hat{a}_k^\dagger |\mathrm{vac}\rangle$ ($k$ integer), which are pulse-like in the continuous-time basis; \textit{i.e.}, $\langle t|k\rangle = g(t-kT)$, where $g(t)$ vanishes for $t \notin (0,T)$, and $\int dt\,g^*(t) g(t) = 1$. On the input side, we take the wavepackets from $k=0$ to $d-1$ as the subspace of interest for quantum communication. The modes are manipulated by EOMs each capable of applying an arbitrary phase shift to each time bin [$\hat{a}_k^\mathrm{(out)} = e^{i\phi_k} \hat{a}_k^\mathrm{(in)}$] and FBGs that implement the tapped delay line [$\hat{a}_n^\mathrm{(out)} = \sum_n c_{n-k} \hat{a}_k^\mathrm{(in)}$], with the values of $c_n$ chosen to ensure unitarity. After such a sequence, the output wavepacket modes described by $\hat{a}_l^\mathrm{(out)}$ are given by
\begin{equation}
\label{e3}
\hat{a}_l^\mathrm{(out)} = \sum_{k=0}^{d-1} W_{lk} \hat{a}_k^\mathrm{(in)}.
\end{equation}
The EOM/FBG network is completely described by the matrix $W$, and, in general, the output includes many possible time-bin modes. For convenience, we also define the projection onto the desired $d$-dimensional output subspace: $V_{lk}=W_{lk}, \;\;\; k,l = 0,1,...,d-1.$. Finally, we consider $d+1$ different parameter sets, so that we label the matrices as $W^{(m)}$ and $V^{(m)}$ ($m=0,1,...,d$).

Ideally, these $d+1$ matrices $V^{(m)}$  will be both unitary and mutually unbiased, in the sense that:
\begin{equation}
\label{e5}
\left| V^{(m)\dagger} V^{(p)} \right|^2 = 
\begin{cases}
\mathbbm{1}_{d\times d}, & m=p \\
\frac{1}{d} \text{ones}(d,d), &  m \neq p,
\end{cases}
\end{equation}
where the modulus-squared is taken element-by-element, $\mathbbm{1}_{d\times d}$ is the $d\times d$ identity matrix, and $\text{ones}(d,d)$ is the $d\times d$ matrix with all elements equal to unity. Equation~(\ref{e5}) is fully equivalent to the standard definition of MUBs in terms of basis-state overlap, by taking each matrix's rows (or columns) as the coefficients of a particular orthonormal basis. If the proposed system can realize a set of such $d+1$ unitary transformations, then it provides a one-to-one mapping for each state from a given MUB to a time-bin eigenstate, permitting direct measurement with a single time-resolving detector.

To demonstrate the ability of our approach to satisfy Eq.~(\ref{e5}) with realistic components, we consider a fixed arrangement consisting of two pairs of EOM/FBG combinations ($N=2$ in Fig.~\ref{fig1}) and examine $d=2,3,4$. The choice of $N=2$ strikes a balance between the number of free parameters and computational efficiency in our numerical optimization.  From an implementation perspective, we must treat the reconfigurability of EOMs and FBG differently. The phase sequence imparted by an EOM can be updated rapidly, on the timescale of the chip rate itself, by simply applying a different radio-frequency voltage pattern. On the other hand, the response of each FBG is essentially fixed, with only slow thermal tuning possible (on the order of seconds).~\cite{Mokhtar2003,Zhang2006,Tian2007} Thus, we impose the constraint that FBG chips remain fixed across all MUBs, whereas the EOM waveforms are free to vary between them.

For a given dimension $d$, we search numerically over all possible phase patterns to minimize the average mean-squared error (MSE) between the actual and desired states, defined as
\begin{equation}
\label{e6}
\begin{split}
\epsilon_\mathrm{MSE} = & \frac{2}{d(d+1)} \sum_{m=0}^{d-1} \sum_{p=m+1}^d  \\ & \times \left\{ \frac{1}{d^2} \sum_{l=0}^{d-1} \sum_{k=0}^{d-1} \left[ \left( V^{(m)\dagger} V^{(p)} \right)_{lk} - \frac{1}{d}    \right]^2 \right\}.
\end{split}
\end{equation}
This provides a measure of closeness to the MUB condition for the $d+1$ configurations (similar to previous ``mubness'' parameters~\cite{Bengtsson2007}). Following MATLAB optimization, we attain solutions for $d=2,3,4$ with $\epsilon_\mathrm{MSE}=1.05\times 10^{-7}, 1.50\times 10^{-4}, 1.60\times 10^{-4}$, respectively.~\cite{Supplemental}

\begin{figure*}[t]
\includegraphics[width=6.75in]{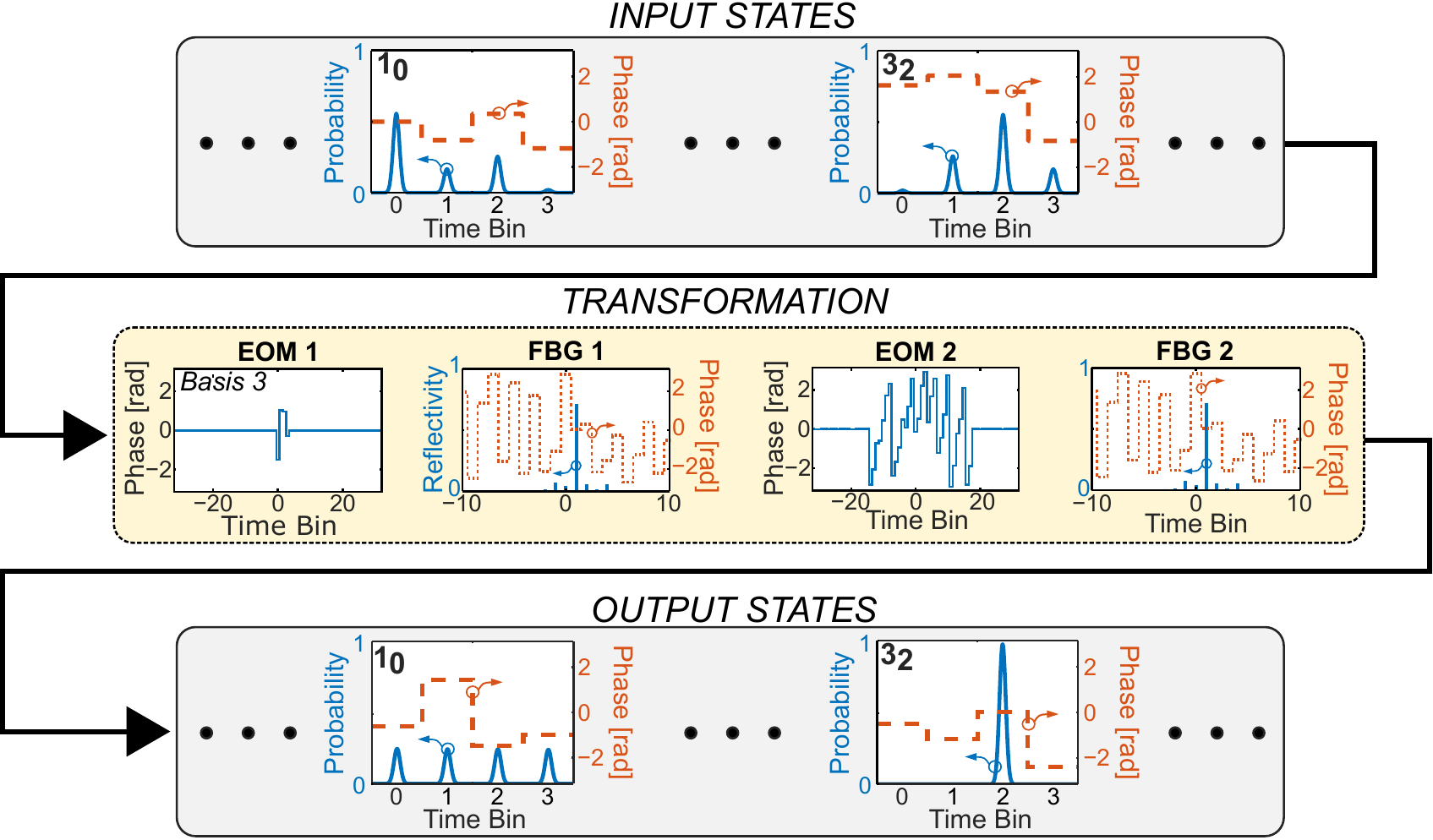}
\caption{Examples for $d=4$. Labels $1_0$ and $3_2$ mark the input/output states corresponding to Alice's choices $|\nu_1[0]\rangle$ and $|\nu_3[2]\rangle$, respectively. The transformation consists of amplitude and phase modulation patterns for basis 3. In practice, the step-function phases need only be constant over the duration of the temporal pulses, with finite transition times acceptable between bins.}
\label{fig2}
\end{figure*}

We analyze each of these solutions in the context of $(d+1)$-basis prepare-and-measure QKD. We assume that the sender, Alice, has at her disposal a source emitting single photons in a superposition of $d$ time bins.  Using a combination of intensity and phase modulators, she imprints onto each wavepacket the appropriate amplitude and phase to match a basis state chosen at random. She then transmits this state to the receiver, Bob, who utilizes the proposed EOM/FBG system to measure the quantum wavepacket in one of the $d+1$ bases, recording the time bin in which the photon is found at the output. Ideally, he will receive a deterministic result whenever his basis matches that from which Alice prepared her state, and a random result for mismatched selections.

Specifically, Alice chooses from the $d(d+1)$ input states
\begin{equation}
\label{e7}
|\nu_m[n]\rangle = \frac{ \sum_{k=0}^{d-1} \left[ V_{nk}^{(m)} \right]^*|k\rangle} { \left( \sum_{l=0}^{d-1} \left| V_{nl}^{(m)} \right|^2  \right)^{1/2} } ; \;\;\; \substack{ m=0,1,...,d \\ n=0,1,...,d-1},
\end{equation}
using the numerically obtained $d+1$ mode transformations $V^{(m)}$ ($m=0,1,...,d$). For ideal unitary matrices, the normalization factor in the denominator is unity, but this is not necessarily the case here, so it is retained. After Alice sends $|\nu_m[n]\rangle$, Bob applies the transformation $V^{(p)}$ and measures the output time bin $q$.

Because Bob post-selects on the photon being found in the $d$-dimensional subspace, we define the detection probability $\mathcal{D}_{pm}[n]$ as the probability that Bob registers a click within bins $0,1,...,d-1$ when Alice prepares $|\nu_m[n]\rangle$ and Bob measures in basis $p$. Similarly, we take $\mathcal{P}_{pm}[q|n]$ as the probability that, conditioned on such a projection into the $d$-dimensional subspace, the input state is measured in bin $q$. Both probabilities are important in evaluating the MUBs: $\mathcal{D}_{pm}[n]$ quantifies the intrinsic efficiency of the measurement process, while $\mathcal{P}_{pm}[q|n]$ the selectivity in distinguishing states.

As an example of how this protocol could be implemented, we illustrate part of the solution for $d=4$ in Fig.~\ref{fig2}. Shown are two of Alice's possible state choices, $|\nu_1 [0]\rangle$ and $|\nu_3 [2]\rangle$; the amplitude varies pulse-to-pulse, unlike MUBs based on phase states.~\cite{Islam2017a,Islam2017b} This is a general feature of our solution method, which places no restrictions on the form of the resultant states, in order to maximize flexibility in the optimization algorithm. In this example, Bob applies the transformation corresponding to basis 3, $V^{(3)}$, which outputs a uniform time-bin superposition for the mismatched choice $|\nu_1 [0]\rangle$, and a single wavepacket (in bin 2) for the matched choice $|\nu_3 [2]\rangle$. 

\begin{figure}
\includegraphics[width=3.4in]{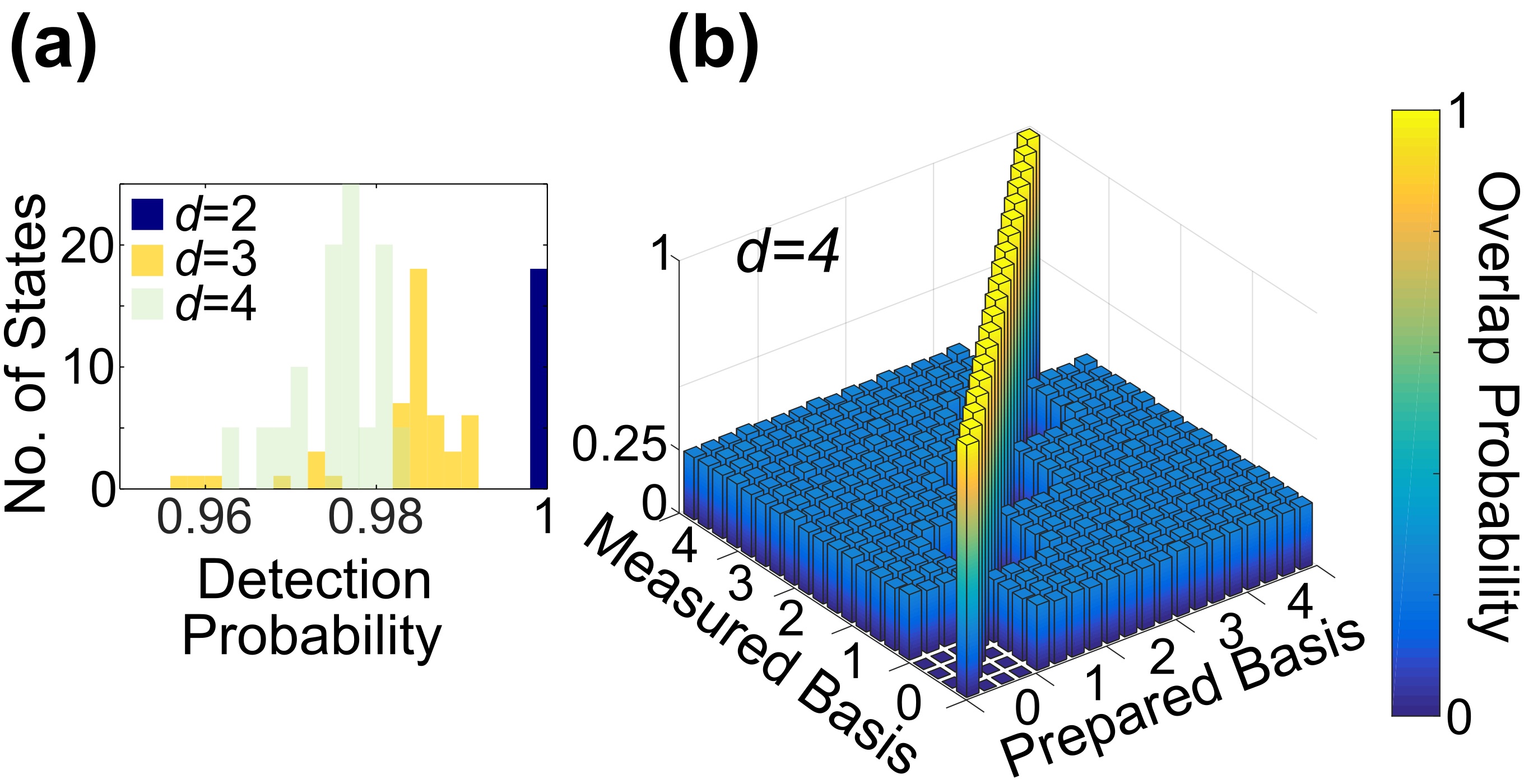}
\caption{Time-bin MUBs. (a) Histogram of detection probabilities for all state/measurement combinations. (b) Post-selected probability distributions for $d=4$.}
\label{fig3}
\end{figure}

In Fig.~\ref{fig3}(a) we provide a histogram of all $d(d+1)^2$ detection probabilities from each solution (the quantities $\mathcal{D}_{pm}[n]$ defined above). Unsurprisingly, given the associated MSE values, $d=2$ has the highest detection probabilities ($>$0.999). Nonetheless, $d=3,4$ still reach values in the range 0.96--0.99: very close to one and substantially higher than the $1/d$ possible with passive DIs.~\cite{Islam2017a} The post-selected probabilities $\mathcal{P}_{pm}[q|n]$ are also near-ideal, as illustrated in Fig.~\ref{fig3}(b), which shows all prepare-and-measure combinations for $d=4$ (Plots of all basis states, measurement settings, and probability distributions for $d=2,3,4$ can be found in the supplemental material.~\cite{Supplemental})

To estimate the impact of physical imperfections on the EOM/FBG measurement, we conduct further simulations that add random errors to the phase of each EOM modulation value and FBG chip (keeping the amplitude unchanged). Each perturbation is drawn independently from a Gaussian distribution of variance $\sigma^2$. To concentrate only on aspects relevant to our proposal, we introduce no other sources of errors (\textit{e.g.}, transmission loss, detector inefficiency, or dark counts). For each phase error $\sigma$, we generate 1,000 mode transformations via Monte Carlo simulation, which we take to represent imperfections in Bob's measurement setup, while Alice continues to send the ideal states. Employing the complete probability distributions from the Monte Carlo simulation [perturbed versions of Fig.~\ref{fig3}(b)], we extract the quantum bit error rate (QBER) and calculate the corresponding secret key fraction (SKF) using the lower bound for an asymmetric $(d+1)$-basis protocol.~\cite{Sheridan2010}

\begin{figure}
\includegraphics[width=3.4in]{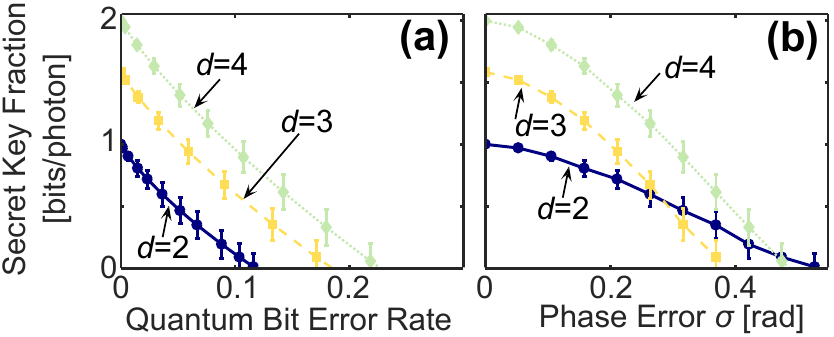}
\caption{Predicted QKD performance with errors. (a) Key fraction against quantum bit error rate. (b) Key fraction against random EOM/FBG phase error $\sigma$.}
\label{fig4}
\end{figure}

In Fig.~\ref{fig4}(a), we show the SKF as a function of QBER, where the error bars indicate statistical uncertainty associated with each data set generated from the Monte Carlo simulation. The behavior matches that expected for high-dimensional QKD: an SKF of $\log_2 d$ at QBER=0, along with greater tolerance to QBER for larger $d$.

In Fig.~\ref{fig4}(b), we plot the SKF against phase error $\sigma$ directly. Though actually the same points as in Fig.~\ref{fig4}(a), by selecting $\sigma$ as the abscissa, we can directly map how phase instabilities degrade QKD performance. Overall, the trend is similar, with the exception that the $d=2$ solution shows greater robustness to phase error at high $\sigma$, with its SKF even exceeding those of $d=3,4$ at $\sigma=0.28,0.47$, respectively. This feature follows from the quality of the solution itself. For as noted above, the metric $\epsilon_\mathrm{MSE}$ is over three orders of magnitude closer to zero for the $d=2$ solution than for $d=3,4$. The results of Fig.~\ref{fig4}(b) then indicate that lower $\epsilon_\mathrm{MSE}$ directly translates to greater robustness against phase error (\textit{i.e.}, lower QBER for a given $\sigma$), validating the utility of the metric $\epsilon_\mathrm{MSE}$ in designing QKD systems based on our proposal.

An important prerequisite toward implementation is matching the characteristic timescales for EOM, FBG, and detector capabilities. With current single-photon detection jitters $<$20 ps,~\cite{Wu2017, Zadeh2018} time-bin spacings must be at least this large; voltage patterns with $\sim$10-ps chips are accessible with modern electronic arbitrary waveform generators,~\cite{Keysight2017} so electro-optic modulation at this speed is achievable. Detailed analysis of our MUB solutions~\cite{Supplemental} implies that $\sim$20 chips would be desirable, giving total code durations $\gtrsim 400$ ps, comfortably within the design space of previously fabricated FBGs.~\cite{Yin2007, Dai2007, Si2010} With thermal tuning,~\cite{Mokhtar2003, Zhang2006, Tian2007} fine adjustments to each chip's phase could compensate for fabrication errors as well. Coupled with picosecond-level synchronization between Alice and Bob via clock recovery of a wavelength-multiplexed reference pulse train,~\cite{Bienfang2004, Patel2012} the necessary timing precision should be attainable even at long distances.

In summary, we have introduced a new scheme for implementing mutually unbiased generation and measurement of time-bin photonic qudits, addressing three persistent challenges in time-bin-based quantum information: (1) preserving single-spatial-mode operation; (2) actualizing multiple MUBs in the same physical setup; and (3) measuring states without $\mathcal{O}(1/d)$ post-selection. While this combination is currently unfulfilled in any other approach, we do note that points (2) and (3) could in principle be addressed by more complex DI networks containing fast phase shifters in each arm and optical switches in lieu of beasmplistters~\cite{Wang2017}---though extremely fast (multi-GHz) switching speed is required. Far more unique is our ability to complete all processing tasks within the innate stability of a single fiber-optic mode [point (1)]: no other single-mode alternative for time-bin MUBs exists at present. And so, in addition to the obvious goal of implementing our new scheme experimentally, we hope our work will motivate the quantum information community to further study the possibilities of coded fiber Bragg gratings for high-dimensional quantum information processing.

\begin{acknowledgments}
We thank B.~Qi and P.~Lougovski for discussions. This work was performed in part at Oak Ridge National Laboratory, operated by UT-Battelle for the U.S. Department of Energy under contract no. DE-AC05-00OR22725. J.M.L. acknowledges support from a Wigner Fellowship at ORNL. N.T.I. and D.J.G. acknowledge support from the ONR MURI program, Wavelength-Agile QKD in a Marine Environment (grant no. N00014-13-1-0627), and the DARPA DSO InPho program. C.C.W.L. acknowledges support from NUS startup grant R-263-000-C78-133/731 and CQT fellow grant.
\end{acknowledgments}

%\bibliography{MUBrefarXiv}

\begin{thebibliography}{31}%
\makeatletter
\providecommand \@ifxundefined [1]{%
 \@ifx{#1\undefined}
}%
\providecommand \@ifnum [1]{%
 \ifnum #1\expandafter \@firstoftwo
 \else \expandafter \@secondoftwo
 \fi
}%
\providecommand \@ifx [1]{%
 \ifx #1\expandafter \@firstoftwo
 \else \expandafter \@secondoftwo
 \fi
}%
\providecommand \natexlab [1]{#1}%
\providecommand \enquote  [1]{``#1''}%
\providecommand \bibnamefont  [1]{#1}%
\providecommand \bibfnamefont [1]{#1}%
\providecommand \citenamefont [1]{#1}%
\providecommand \href@noop [0]{\@secondoftwo}%
\providecommand \href [0]{\begingroup \@sanitize@url \@href}%
\providecommand \@href[1]{\@@startlink{#1}\@@href}%
\providecommand \@@href[1]{\endgroup#1\@@endlink}%
\providecommand \@sanitize@url [0]{\catcode `\\12\catcode `\$12\catcode
  `\&12\catcode `\#12\catcode `\^12\catcode `\_12\catcode `\%12\relax}%
\providecommand \@@startlink[1]{}%
\providecommand \@@endlink[0]{}%
\providecommand \url  [0]{\begingroup\@sanitize@url \@url }%
\providecommand \@url [1]{\endgroup\@href {#1}{\urlprefix }}%
\providecommand \urlprefix  [0]{URL }%
\providecommand \Eprint [0]{\href }%
\providecommand \doibase [0]{http://dx.doi.org/}%
\providecommand \selectlanguage [0]{\@gobble}%
\providecommand \bibinfo  [0]{\@secondoftwo}%
\providecommand \bibfield  [0]{\@secondoftwo}%
\providecommand \translation [1]{[#1]}%
\providecommand \BibitemOpen [0]{}%
\providecommand \bibitemStop [0]{}%
\providecommand \bibitemNoStop [0]{.\EOS\space}%
\providecommand \EOS [0]{\spacefactor3000\relax}%
\providecommand \BibitemShut  [1]{\csname bibitem#1\endcsname}%
\let\auto@bib@innerbib\@empty
%</preamble>
\bibitem [{\citenamefont {Wootters}\ and\ \citenamefont
  {Fields}(1989)}]{Wootters1989}%
  \BibitemOpen
  \bibfield  {author} {\bibinfo {author} {\bibfnamefont {W.~K.}\ \bibnamefont
  {Wootters}}\ and\ \bibinfo {author} {\bibfnamefont {B.~D.}\ \bibnamefont
  {Fields}},\ }\href {\doibase http://dx.doi.org/10.1016/0003-4916(89)90322-9}
  {\bibfield  {journal} {\bibinfo  {journal} {Ann. Phys.}\ }\textbf {\bibinfo
  {volume} {191}},\ \bibinfo {pages} {363} (\bibinfo {year}
  {1989})}\BibitemShut {NoStop}%
\bibitem [{\citenamefont {Durt}\ \emph {et~al.}(2010)\citenamefont {Durt},
  \citenamefont {Englert}, \citenamefont {Bengtsson},\ and\ \citenamefont
  {\.{Z}yczkowski}}]{Durt2010}%
  \BibitemOpen
  \bibfield  {author} {\bibinfo {author} {\bibfnamefont {T.}~\bibnamefont
  {Durt}}, \bibinfo {author} {\bibfnamefont {B.-G.}\ \bibnamefont {Englert}},
  \bibinfo {author} {\bibfnamefont {I.}~\bibnamefont {Bengtsson}}, \ and\
  \bibinfo {author} {\bibfnamefont {K.}~\bibnamefont {\.{Z}yczkowski}},\
  }\href@noop {} {\bibfield  {journal} {\bibinfo  {journal} {Int. J. Quantum
  Inf.}\ }\textbf {\bibinfo {volume} {8}},\ \bibinfo {pages} {535} (\bibinfo
  {year} {2010})}\BibitemShut {NoStop}%
\bibitem [{\citenamefont {Adamson}\ and\ \citenamefont
  {Steinberg}(2010)}]{Adamson2010}%
  \BibitemOpen
  \bibfield  {author} {\bibinfo {author} {\bibfnamefont {R.~B.~A.}\
  \bibnamefont {Adamson}}\ and\ \bibinfo {author} {\bibfnamefont {A.~M.}\
  \bibnamefont {Steinberg}},\ }\href {\doibase 10.1103/PhysRevLett.105.030406}
  {\bibfield  {journal} {\bibinfo  {journal} {Phys. Rev. Lett.}\ }\textbf
  {\bibinfo {volume} {105}},\ \bibinfo {pages} {030406} (\bibinfo {year}
  {2010})}\BibitemShut {NoStop}%
\bibitem [{\citenamefont {Gisin}\ \emph {et~al.}(2002)\citenamefont {Gisin},
  \citenamefont {Ribordy}, \citenamefont {Tittel},\ and\ \citenamefont
  {Zbinden}}]{Gisin2002}%
  \BibitemOpen
  \bibfield  {author} {\bibinfo {author} {\bibfnamefont {N.}~\bibnamefont
  {Gisin}}, \bibinfo {author} {\bibfnamefont {G.}~\bibnamefont {Ribordy}},
  \bibinfo {author} {\bibfnamefont {W.}~\bibnamefont {Tittel}}, \ and\ \bibinfo
  {author} {\bibfnamefont {H.}~\bibnamefont {Zbinden}},\ }\href {\doibase
  10.1103/RevModPhys.74.145} {\bibfield  {journal} {\bibinfo  {journal} {Rev.
  Mod. Phys.}\ }\textbf {\bibinfo {volume} {74}},\ \bibinfo {pages} {145}
  (\bibinfo {year} {2002})}\BibitemShut {NoStop}%
\bibitem [{\citenamefont {Scarani}\ \emph {et~al.}(2009)\citenamefont
  {Scarani}, \citenamefont {Bechmann-Pasquinucci}, \citenamefont {Cerf},
  \citenamefont {Du\ifmmode~\check{s}\else \v{s}\fi{}ek}, \citenamefont
  {L\"utkenhaus},\ and\ \citenamefont {Peev}}]{Scarani2009}%
  \BibitemOpen
  \bibfield  {author} {\bibinfo {author} {\bibfnamefont {V.}~\bibnamefont
  {Scarani}}, \bibinfo {author} {\bibfnamefont {H.}~\bibnamefont
  {Bechmann-Pasquinucci}}, \bibinfo {author} {\bibfnamefont {N.~J.}\
  \bibnamefont {Cerf}}, \bibinfo {author} {\bibfnamefont {M.}~\bibnamefont
  {Du\ifmmode~\check{s}\else \v{s}\fi{}ek}}, \bibinfo {author} {\bibfnamefont
  {N.}~\bibnamefont {L\"utkenhaus}}, \ and\ \bibinfo {author} {\bibfnamefont
  {M.}~\bibnamefont {Peev}},\ }\href {\doibase 10.1103/RevModPhys.81.1301}
  {\bibfield  {journal} {\bibinfo  {journal} {Rev. Mod. Phys.}\ }\textbf
  {\bibinfo {volume} {81}},\ \bibinfo {pages} {1301} (\bibinfo {year}
  {2009})}\BibitemShut {NoStop}%
\bibitem [{\citenamefont {Brougham}\ \emph {et~al.}(2013)\citenamefont
  {Brougham}, \citenamefont {Barnett}, \citenamefont {McCusker}, \citenamefont
  {Kwiat},\ and\ \citenamefont {Gauthier}}]{Brougham2013a}%
  \BibitemOpen
  \bibfield  {author} {\bibinfo {author} {\bibfnamefont {T.}~\bibnamefont
  {Brougham}}, \bibinfo {author} {\bibfnamefont {S.~M.}\ \bibnamefont
  {Barnett}}, \bibinfo {author} {\bibfnamefont {K.~T.}\ \bibnamefont
  {McCusker}}, \bibinfo {author} {\bibfnamefont {P.~G.}\ \bibnamefont {Kwiat}},
  \ and\ \bibinfo {author} {\bibfnamefont {D.~J.}\ \bibnamefont {Gauthier}},\
  }\href {http://stacks.iop.org/0953-4075/46/i=10/a=104010} {\bibfield
  {journal} {\bibinfo  {journal} {J. Phys. B}\ }\textbf {\bibinfo {volume}
  {46}},\ \bibinfo {pages} {104010} (\bibinfo {year} {2013})}\BibitemShut
  {NoStop}%
\bibitem [{\citenamefont {Islam}\ \emph
  {et~al.}(2017{\natexlab{a}})\citenamefont {Islam}, \citenamefont {Cahall},
  \citenamefont {Aragoneses}, \citenamefont {Lezama}, \citenamefont {Kim},\
  and\ \citenamefont {Gauthier}}]{Islam2017a}%
  \BibitemOpen
  \bibfield  {author} {\bibinfo {author} {\bibfnamefont {N.~T.}\ \bibnamefont
  {Islam}}, \bibinfo {author} {\bibfnamefont {C.}~\bibnamefont {Cahall}},
  \bibinfo {author} {\bibfnamefont {A.}~\bibnamefont {Aragoneses}}, \bibinfo
  {author} {\bibfnamefont {A.}~\bibnamefont {Lezama}}, \bibinfo {author}
  {\bibfnamefont {J.}~\bibnamefont {Kim}}, \ and\ \bibinfo {author}
  {\bibfnamefont {D.~J.}\ \bibnamefont {Gauthier}},\ }\href {\doibase
  10.1103/PhysRevApplied.7.044010} {\bibfield  {journal} {\bibinfo  {journal}
  {Phys. Rev. Applied}\ }\textbf {\bibinfo {volume} {7}},\ \bibinfo {pages}
  {044010} (\bibinfo {year} {2017}{\natexlab{a}})}\BibitemShut {NoStop}%
\bibitem [{\citenamefont {Islam}\ \emph
  {et~al.}(2017{\natexlab{b}})\citenamefont {Islam}, \citenamefont {Lim},
  \citenamefont {Cahall}, \citenamefont {Kim},\ and\ \citenamefont
  {Gauthier}}]{Islam2017b}%
  \BibitemOpen
  \bibfield  {author} {\bibinfo {author} {\bibfnamefont {N.~T.}\ \bibnamefont
  {Islam}}, \bibinfo {author} {\bibfnamefont {C.~C.~W.}\ \bibnamefont {Lim}},
  \bibinfo {author} {\bibfnamefont {C.}~\bibnamefont {Cahall}}, \bibinfo
  {author} {\bibfnamefont {J.}~\bibnamefont {Kim}}, \ and\ \bibinfo {author}
  {\bibfnamefont {D.~J.}\ \bibnamefont {Gauthier}},\ }\href {\doibase
  10.1126/sciadv.1701491} {\bibfield  {journal} {\bibinfo  {journal} {Sci.
  Adv.}\ }\textbf {\bibinfo {volume} {3}},\ \bibinfo {pages} {e1701491}
  (\bibinfo {year} {2017}{\natexlab{b}})}\BibitemShut {NoStop}%
\bibitem [{\citenamefont {Hillerkuss}\ \emph {et~al.}(2010)\citenamefont
  {Hillerkuss}, \citenamefont {Winter}, \citenamefont {Teschke}, \citenamefont
  {Marculescu}, \citenamefont {Li}, \citenamefont {Sigurdsson}, \citenamefont
  {Worms}, \citenamefont {Ezra}, \citenamefont {Narkiss}, \citenamefont
  {Freude},\ and\ \citenamefont {Leuthold}}]{Hillerkuss2010}%
  \BibitemOpen
  \bibfield  {author} {\bibinfo {author} {\bibfnamefont {D.}~\bibnamefont
  {Hillerkuss}}, \bibinfo {author} {\bibfnamefont {M.}~\bibnamefont {Winter}},
  \bibinfo {author} {\bibfnamefont {M.}~\bibnamefont {Teschke}}, \bibinfo
  {author} {\bibfnamefont {A.}~\bibnamefont {Marculescu}}, \bibinfo {author}
  {\bibfnamefont {J.}~\bibnamefont {Li}}, \bibinfo {author} {\bibfnamefont
  {G.}~\bibnamefont {Sigurdsson}}, \bibinfo {author} {\bibfnamefont
  {K.}~\bibnamefont {Worms}}, \bibinfo {author} {\bibfnamefont {S.~B.}\
  \bibnamefont {Ezra}}, \bibinfo {author} {\bibfnamefont {N.}~\bibnamefont
  {Narkiss}}, \bibinfo {author} {\bibfnamefont {W.}~\bibnamefont {Freude}}, \
  and\ \bibinfo {author} {\bibfnamefont {J.}~\bibnamefont {Leuthold}},\ }\href
  {\doibase 10.1364/OE.18.009324} {\bibfield  {journal} {\bibinfo  {journal}
  {Opt. Express}\ }\textbf {\bibinfo {volume} {18}},\ \bibinfo {pages} {9324}
  (\bibinfo {year} {2010})}\BibitemShut {NoStop}%
\bibitem [{\citenamefont {Hillerkuss}\ \emph {et~al.}(2011)\citenamefont
  {Hillerkuss}, \citenamefont {Schmogrow}, \citenamefont {Schellinger},
  \citenamefont {Jordan}, \citenamefont {Winter}, \citenamefont {Huber},
  \citenamefont {Vallaitis}, \citenamefont {Bonk}, \citenamefont {Kleinow},
  \citenamefont {Frey}, \citenamefont {Roeger}, \citenamefont {Koenig},
  \citenamefont {Ludwig}, \citenamefont {Marculescu}, \citenamefont {Li},
  \citenamefont {Hoh}, \citenamefont {Dreschmann}, \citenamefont {Meyer},
  \citenamefont {Huebner}, \citenamefont {Becker}, \citenamefont {Koos},
  \citenamefont {Freude},\ and\ \citenamefont {Leuthold}}]{Hillerkuss2011}%
  \BibitemOpen
  \bibfield  {author} {\bibinfo {author} {\bibfnamefont {D.}~\bibnamefont
  {Hillerkuss}}, \bibinfo {author} {\bibfnamefont {R.}~\bibnamefont
  {Schmogrow}}, \bibinfo {author} {\bibfnamefont {T.}~\bibnamefont
  {Schellinger}}, \bibinfo {author} {\bibfnamefont {M.}~\bibnamefont {Jordan}},
  \bibinfo {author} {\bibfnamefont {M.}~\bibnamefont {Winter}}, \bibinfo
  {author} {\bibfnamefont {G.}~\bibnamefont {Huber}}, \bibinfo {author}
  {\bibfnamefont {T.}~\bibnamefont {Vallaitis}}, \bibinfo {author}
  {\bibfnamefont {R.}~\bibnamefont {Bonk}}, \bibinfo {author} {\bibfnamefont
  {P.}~\bibnamefont {Kleinow}}, \bibinfo {author} {\bibfnamefont
  {F.}~\bibnamefont {Frey}}, \bibinfo {author} {\bibfnamefont {M.}~\bibnamefont
  {Roeger}}, \bibinfo {author} {\bibfnamefont {S.}~\bibnamefont {Koenig}},
  \bibinfo {author} {\bibfnamefont {A.}~\bibnamefont {Ludwig}}, \bibinfo
  {author} {\bibfnamefont {A.}~\bibnamefont {Marculescu}}, \bibinfo {author}
  {\bibfnamefont {J.}~\bibnamefont {Li}}, \bibinfo {author} {\bibfnamefont
  {M.}~\bibnamefont {Hoh}}, \bibinfo {author} {\bibfnamefont {M.}~\bibnamefont
  {Dreschmann}}, \bibinfo {author} {\bibfnamefont {J.}~\bibnamefont {Meyer}},
  \bibinfo {author} {\bibfnamefont {M.}~\bibnamefont {Huebner}}, \bibinfo
  {author} {\bibfnamefont {J.}~\bibnamefont {Becker}}, \bibinfo {author}
  {\bibfnamefont {C.}~\bibnamefont {Koos}}, \bibinfo {author} {\bibfnamefont
  {W.}~\bibnamefont {Freude}}, \ and\ \bibinfo {author} {\bibfnamefont
  {J.}~\bibnamefont {Leuthold}},\ }\href@noop {} {\bibfield  {journal}
  {\bibinfo  {journal} {Nat. Photon.}\ }\textbf {\bibinfo {volume} {5}},\
  \bibinfo {pages} {364} (\bibinfo {year} {2011})}\BibitemShut {NoStop}%
\bibitem [{\citenamefont {Cerf}\ \emph {et~al.}(2002)\citenamefont {Cerf},
  \citenamefont {Bourennane}, \citenamefont {Karlsson},\ and\ \citenamefont
  {Gisin}}]{Cerf2002}%
  \BibitemOpen
  \bibfield  {author} {\bibinfo {author} {\bibfnamefont {N.~J.}\ \bibnamefont
  {Cerf}}, \bibinfo {author} {\bibfnamefont {M.}~\bibnamefont {Bourennane}},
  \bibinfo {author} {\bibfnamefont {A.}~\bibnamefont {Karlsson}}, \ and\
  \bibinfo {author} {\bibfnamefont {N.}~\bibnamefont {Gisin}},\ }\href
  {\doibase 10.1103/PhysRevLett.88.127902} {\bibfield  {journal} {\bibinfo
  {journal} {Phys. Rev. Lett.}\ }\textbf {\bibinfo {volume} {88}},\ \bibinfo
  {pages} {127902} (\bibinfo {year} {2002})}\BibitemShut {NoStop}%
\bibitem [{\citenamefont {Sheridan}\ and\ \citenamefont
  {Scarani}(2010)}]{Sheridan2010}%
  \BibitemOpen
  \bibfield  {author} {\bibinfo {author} {\bibfnamefont {L.}~\bibnamefont
  {Sheridan}}\ and\ \bibinfo {author} {\bibfnamefont {V.}~\bibnamefont
  {Scarani}},\ }\href {\doibase 10.1103/PhysRevA.82.030301} {\bibfield
  {journal} {\bibinfo  {journal} {Phys. Rev. A}\ }\textbf {\bibinfo {volume}
  {82}},\ \bibinfo {pages} {030301} (\bibinfo {year} {2010})}\BibitemShut
  {NoStop}%
\bibitem [{\citenamefont {Lima}\ \emph {et~al.}(2011)\citenamefont {Lima},
  \citenamefont {Neves}, \citenamefont {Guzm\'{a}n}, \citenamefont {G\'{o}mez},
  \citenamefont {Nogueira}, \citenamefont {Delgado}, \citenamefont {Vargas},\
  and\ \citenamefont {Saavedra}}]{Lima2011}%
  \BibitemOpen
  \bibfield  {author} {\bibinfo {author} {\bibfnamefont {G.}~\bibnamefont
  {Lima}}, \bibinfo {author} {\bibfnamefont {L.}~\bibnamefont {Neves}},
  \bibinfo {author} {\bibfnamefont {R.}~\bibnamefont {Guzm\'{a}n}}, \bibinfo
  {author} {\bibfnamefont {E.~S.}\ \bibnamefont {G\'{o}mez}}, \bibinfo {author}
  {\bibfnamefont {W.~A.~T.}\ \bibnamefont {Nogueira}}, \bibinfo {author}
  {\bibfnamefont {A.}~\bibnamefont {Delgado}}, \bibinfo {author} {\bibfnamefont
  {A.}~\bibnamefont {Vargas}}, \ and\ \bibinfo {author} {\bibfnamefont
  {C.}~\bibnamefont {Saavedra}},\ }\href {\doibase 10.1364/OE.19.003542}
  {\bibfield  {journal} {\bibinfo  {journal} {Opt. Express}\ }\textbf {\bibinfo
  {volume} {19}},\ \bibinfo {pages} {3542} (\bibinfo {year}
  {2011})}\BibitemShut {NoStop}%
\bibitem [{\citenamefont {Giovannini}\ \emph {et~al.}(2013)\citenamefont
  {Giovannini}, \citenamefont {Romero}, \citenamefont {Leach}, \citenamefont
  {Dudley}, \citenamefont {Forbes},\ and\ \citenamefont
  {Padgett}}]{Giovannini2013}%
  \BibitemOpen
  \bibfield  {author} {\bibinfo {author} {\bibfnamefont {D.}~\bibnamefont
  {Giovannini}}, \bibinfo {author} {\bibfnamefont {J.}~\bibnamefont {Romero}},
  \bibinfo {author} {\bibfnamefont {J.}~\bibnamefont {Leach}}, \bibinfo
  {author} {\bibfnamefont {A.}~\bibnamefont {Dudley}}, \bibinfo {author}
  {\bibfnamefont {A.}~\bibnamefont {Forbes}}, \ and\ \bibinfo {author}
  {\bibfnamefont {M.~J.}\ \bibnamefont {Padgett}},\ }\href {\doibase
  10.1103/PhysRevLett.110.143601} {\bibfield  {journal} {\bibinfo  {journal}
  {Phys. Rev. Lett.}\ }\textbf {\bibinfo {volume} {110}},\ \bibinfo {pages}
  {143601} (\bibinfo {year} {2013})}\BibitemShut {NoStop}%
\bibitem [{\citenamefont {Yin}\ and\ \citenamefont
  {Richardson}(2007)}]{Yin2007}%
  \BibitemOpen
  \bibfield  {author} {\bibinfo {author} {\bibfnamefont {H.}~\bibnamefont
  {Yin}}\ and\ \bibinfo {author} {\bibfnamefont {D.~J.}\ \bibnamefont
  {Richardson}},\ }\href@noop {} {\emph {\bibinfo {title} {Optical Code
  Division Multiple Access Communication Networks}}}\ (\bibinfo  {publisher}
  {Springer},\ \bibinfo {year} {2007})\BibitemShut {NoStop}%
\bibitem [{\citenamefont {Dai}\ \emph {et~al.}(2007)\citenamefont {Dai},
  \citenamefont {Chen}, \citenamefont {Zhang}, \citenamefont {Sun},\ and\
  \citenamefont {Xie}}]{Dai2007}%
  \BibitemOpen
  \bibfield  {author} {\bibinfo {author} {\bibfnamefont {Y.}~\bibnamefont
  {Dai}}, \bibinfo {author} {\bibfnamefont {X.}~\bibnamefont {Chen}}, \bibinfo
  {author} {\bibfnamefont {Y.}~\bibnamefont {Zhang}}, \bibinfo {author}
  {\bibfnamefont {J.}~\bibnamefont {Sun}}, \ and\ \bibinfo {author}
  {\bibfnamefont {S.}~\bibnamefont {Xie}},\ }in\ \href@noop {} {\emph {\bibinfo
  {booktitle} {Optical Fiber Communication Conference}}}\ (\bibinfo
  {organization} {Optical Society of America},\ \bibinfo {year} {2007})\ p.\
  \bibinfo {pages} {JWA28}\BibitemShut {NoStop}%
\bibitem [{\citenamefont {Chen}, \citenamefont {Chen},\ and\ \citenamefont
  {Xie}(2009)}]{Chen2009}%
  \BibitemOpen
  \bibfield  {author} {\bibinfo {author} {\bibfnamefont {H.}~\bibnamefont
  {Chen}}, \bibinfo {author} {\bibfnamefont {M.}~\bibnamefont {Chen}}, \ and\
  \bibinfo {author} {\bibfnamefont {S.}~\bibnamefont {Xie}},\ }\href
  {http://jlt.osa.org/abstract.cfm?URI=jlt-27-21-4848} {\bibfield  {journal}
  {\bibinfo  {journal} {J. Lightw. Technol.}\ }\textbf {\bibinfo {volume}
  {27}},\ \bibinfo {pages} {4848} (\bibinfo {year} {2009})}\BibitemShut
  {NoStop}%
\bibitem [{\citenamefont {Wooten}\ \emph {et~al.}(2000)\citenamefont {Wooten},
  \citenamefont {Kissa}, \citenamefont {Yi-Yan}, \citenamefont {Murphy},
  \citenamefont {Lafaw}, \citenamefont {Hallemeier}, \citenamefont {Maack},
  \citenamefont {Attanasio}, \citenamefont {Fritz}, \citenamefont {McBrien},\
  and\ \citenamefont {Bossi}}]{Wooten2000}%
  \BibitemOpen
  \bibfield  {author} {\bibinfo {author} {\bibfnamefont {E.~L.}\ \bibnamefont
  {Wooten}}, \bibinfo {author} {\bibfnamefont {K.~M.}\ \bibnamefont {Kissa}},
  \bibinfo {author} {\bibfnamefont {A.}~\bibnamefont {Yi-Yan}}, \bibinfo
  {author} {\bibfnamefont {E.~J.}\ \bibnamefont {Murphy}}, \bibinfo {author}
  {\bibfnamefont {D.~A.}\ \bibnamefont {Lafaw}}, \bibinfo {author}
  {\bibfnamefont {P.~F.}\ \bibnamefont {Hallemeier}}, \bibinfo {author}
  {\bibfnamefont {D.}~\bibnamefont {Maack}}, \bibinfo {author} {\bibfnamefont
  {D.~V.}\ \bibnamefont {Attanasio}}, \bibinfo {author} {\bibfnamefont {D.~J.}\
  \bibnamefont {Fritz}}, \bibinfo {author} {\bibfnamefont {G.~J.}\ \bibnamefont
  {McBrien}}, \ and\ \bibinfo {author} {\bibfnamefont {D.~E.}\ \bibnamefont
  {Bossi}},\ }\href {\doibase 10.1109/2944.826874} {\bibfield  {journal}
  {\bibinfo  {journal} {IEEE J. Sel. Top. Quantum Electron.}\ }\textbf
  {\bibinfo {volume} {6}},\ \bibinfo {pages} {69} (\bibinfo {year}
  {2000})}\BibitemShut {NoStop}%
\bibitem [{\citenamefont {Lukens}\ and\ \citenamefont
  {Lougovski}(2017)}]{Lukens2017}%
  \BibitemOpen
  \bibfield  {author} {\bibinfo {author} {\bibfnamefont {J.~M.}\ \bibnamefont
  {Lukens}}\ and\ \bibinfo {author} {\bibfnamefont {P.}~\bibnamefont
  {Lougovski}},\ }\href {\doibase 10.1364/OPTICA.4.000008} {\bibfield
  {journal} {\bibinfo  {journal} {Optica}\ }\textbf {\bibinfo {volume} {4}},\
  \bibinfo {pages} {8} (\bibinfo {year} {2017})}\BibitemShut {NoStop}%
\bibitem [{\citenamefont {Mokhtar}\ \emph {et~al.}(2003)\citenamefont
  {Mokhtar}, \citenamefont {Ibsen}, \citenamefont {Teh},\ and\ \citenamefont
  {Richardson}}]{Mokhtar2003}%
  \BibitemOpen
  \bibfield  {author} {\bibinfo {author} {\bibfnamefont {M.~R.}\ \bibnamefont
  {Mokhtar}}, \bibinfo {author} {\bibfnamefont {M.}~\bibnamefont {Ibsen}},
  \bibinfo {author} {\bibfnamefont {P.~C.}\ \bibnamefont {Teh}}, \ and\
  \bibinfo {author} {\bibfnamefont {D.~J.}\ \bibnamefont {Richardson}},\ }\href
  {\doibase 10.1109/LPT.2003.807902} {\bibfield  {journal} {\bibinfo  {journal}
  {IEEE Photon. Technol. Lett.}\ }\textbf {\bibinfo {volume} {15}},\ \bibinfo
  {pages} {431} (\bibinfo {year} {2003})}\BibitemShut {NoStop}%
\bibitem [{\citenamefont {Zhang}\ \emph {et~al.}(2006)\citenamefont {Zhang},
  \citenamefont {Tian}, \citenamefont {Mokhtar}, \citenamefont {Petropoulos},
  \citenamefont {Richardson},\ and\ \citenamefont {Ibsen}}]{Zhang2006}%
  \BibitemOpen
  \bibfield  {author} {\bibinfo {author} {\bibfnamefont {Z.}~\bibnamefont
  {Zhang}}, \bibinfo {author} {\bibfnamefont {C.}~\bibnamefont {Tian}},
  \bibinfo {author} {\bibfnamefont {M.~R.}\ \bibnamefont {Mokhtar}}, \bibinfo
  {author} {\bibfnamefont {P.}~\bibnamefont {Petropoulos}}, \bibinfo {author}
  {\bibfnamefont {D.~J.}\ \bibnamefont {Richardson}}, \ and\ \bibinfo {author}
  {\bibfnamefont {M.}~\bibnamefont {Ibsen}},\ }\href {\doibase
  10.1109/LPT.2006.875062} {\bibfield  {journal} {\bibinfo  {journal} {IEEE
  Photon. Technol. Lett.}\ }\textbf {\bibinfo {volume} {18}},\ \bibinfo {pages}
  {1216} (\bibinfo {year} {2006})}\BibitemShut {NoStop}%
\bibitem [{\citenamefont {Tian}\ \emph {et~al.}(2007)\citenamefont {Tian},
  \citenamefont {Zhang}, \citenamefont {Ibsen}, \citenamefont {Petropoulos},\
  and\ \citenamefont {Richardson}}]{Tian2007}%
  \BibitemOpen
  \bibfield  {author} {\bibinfo {author} {\bibfnamefont {C.}~\bibnamefont
  {Tian}}, \bibinfo {author} {\bibfnamefont {Z.}~\bibnamefont {Zhang}},
  \bibinfo {author} {\bibfnamefont {M.}~\bibnamefont {Ibsen}}, \bibinfo
  {author} {\bibfnamefont {P.}~\bibnamefont {Petropoulos}}, \ and\ \bibinfo
  {author} {\bibfnamefont {D.~J.}\ \bibnamefont {Richardson}},\ }\href
  {\doibase 10.1109/JSTQE.2007.897668} {\bibfield  {journal} {\bibinfo
  {journal} {IEEE J. Sel. Top. Quantum Electron.}\ }\textbf {\bibinfo {volume}
  {13}},\ \bibinfo {pages} {1480} (\bibinfo {year} {2007})}\BibitemShut
  {NoStop}%
\bibitem [{\citenamefont {Bengtsson}(2007)}]{Bengtsson2007}%
  \BibitemOpen
  \bibfield  {author} {\bibinfo {author} {\bibfnamefont {I.}~\bibnamefont
  {Bengtsson}},\ }\href {\doibase 10.1063/1.2713445} {\bibfield  {journal}
  {\bibinfo  {journal} {AIP Conference Proceedings}\ }\textbf {\bibinfo
  {volume} {889}},\ \bibinfo {pages} {40} (\bibinfo {year} {2007})}\BibitemShut
  {NoStop}%
\bibitem [{Sup()}]{Supplemental}%
  \BibitemOpen
  \href@noop {} {}\bibinfo {note} {See supplemental material on p. 6 for
  details on the specific solutions and their properties.}\BibitemShut {Stop}%
\bibitem [{\citenamefont {Wu}\ \emph {et~al.}(2017)\citenamefont {Wu},
  \citenamefont {You}, \citenamefont {Chen}, \citenamefont {Li}, \citenamefont
  {He}, \citenamefont {Lv}, \citenamefont {Wang},\ and\ \citenamefont
  {Xie}}]{Wu2017}%
  \BibitemOpen
  \bibfield  {author} {\bibinfo {author} {\bibfnamefont {J.}~\bibnamefont
  {Wu}}, \bibinfo {author} {\bibfnamefont {L.}~\bibnamefont {You}}, \bibinfo
  {author} {\bibfnamefont {S.}~\bibnamefont {Chen}}, \bibinfo {author}
  {\bibfnamefont {H.}~\bibnamefont {Li}}, \bibinfo {author} {\bibfnamefont
  {Y.}~\bibnamefont {He}}, \bibinfo {author} {\bibfnamefont {C.}~\bibnamefont
  {Lv}}, \bibinfo {author} {\bibfnamefont {Z.}~\bibnamefont {Wang}}, \ and\
  \bibinfo {author} {\bibfnamefont {X.}~\bibnamefont {Xie}},\ }\href {\doibase
  10.1364/AO.56.002195} {\bibfield  {journal} {\bibinfo  {journal} {Appl.
  Opt.}\ }\textbf {\bibinfo {volume} {56}},\ \bibinfo {pages} {2195} (\bibinfo
  {year} {2017})}\BibitemShut {NoStop}%
\bibitem [{\citenamefont {Zadeh}\ \emph {et~al.}(2018)\citenamefont {Zadeh},
  \citenamefont {Los}, \citenamefont {Gourgues}, \citenamefont {Bulgarini},
  \citenamefont {Dobrovolskiy}, \citenamefont {Zwiller},\ and\ \citenamefont
  {Dorenbos}}]{Zadeh2018}%
  \BibitemOpen
  \bibfield  {author} {\bibinfo {author} {\bibfnamefont {I.~E.}\ \bibnamefont
  {Zadeh}}, \bibinfo {author} {\bibfnamefont {J.~W.~N.}\ \bibnamefont {Los}},
  \bibinfo {author} {\bibfnamefont {R.~B.~M.}\ \bibnamefont {Gourgues}},
  \bibinfo {author} {\bibfnamefont {G.}~\bibnamefont {Bulgarini}}, \bibinfo
  {author} {\bibfnamefont {S.~M.}\ \bibnamefont {Dobrovolskiy}}, \bibinfo
  {author} {\bibfnamefont {V.}~\bibnamefont {Zwiller}}, \ and\ \bibinfo
  {author} {\bibfnamefont {S.~N.}\ \bibnamefont {Dorenbos}},\ }\href@noop {}
  {\bibfield  {journal} {\bibinfo  {journal} {arXiv:1801.06574}\ } (\bibinfo
  {year} {2018})}\BibitemShut {NoStop}%
\bibitem [{\citenamefont {{Keysight Technologies}}(2017)}]{Keysight2017}%
  \BibitemOpen
  \bibfield  {author} {\bibinfo {author} {\bibnamefont {{Keysight
  Technologies}}},\ }\href@noop {} {\enquote {\bibinfo {title} {{AXIe}
  aribitrary waveform generators},}\ }\bibinfo {howpublished}
  {\url{www.keysight.com}} (\bibinfo {year} {2017})\BibitemShut {NoStop}%
\bibitem [{\citenamefont {Si}\ \emph {et~al.}(2010)\citenamefont {Si},
  \citenamefont {Yin}, \citenamefont {Xin}, \citenamefont {Chen}, \citenamefont
  {Chen},\ and\ \citenamefont {Xie}}]{Si2010}%
  \BibitemOpen
  \bibfield  {author} {\bibinfo {author} {\bibfnamefont {Z.}~\bibnamefont
  {Si}}, \bibinfo {author} {\bibfnamefont {F.}~\bibnamefont {Yin}}, \bibinfo
  {author} {\bibfnamefont {M.}~\bibnamefont {Xin}}, \bibinfo {author}
  {\bibfnamefont {H.}~\bibnamefont {Chen}}, \bibinfo {author} {\bibfnamefont
  {M.}~\bibnamefont {Chen}}, \ and\ \bibinfo {author} {\bibfnamefont
  {S.}~\bibnamefont {Xie}},\ }\href {\doibase 10.1364/OL.35.000229} {\bibfield
  {journal} {\bibinfo  {journal} {Opt. Lett.}\ }\textbf {\bibinfo {volume}
  {35}},\ \bibinfo {pages} {229} (\bibinfo {year} {2010})}\BibitemShut
  {NoStop}%
\bibitem [{\citenamefont {Bienfang}\ \emph {et~al.}(2004)\citenamefont
  {Bienfang}, \citenamefont {Gross}, \citenamefont {Mink}, \citenamefont
  {Hershman}, \citenamefont {Nakassis}, \citenamefont {Tang}, \citenamefont
  {Lu}, \citenamefont {Su}, \citenamefont {Clark}, \citenamefont {Williams},
  \citenamefont {Hagley},\ and\ \citenamefont {Wen}}]{Bienfang2004}%
  \BibitemOpen
  \bibfield  {author} {\bibinfo {author} {\bibfnamefont {J.~C.}\ \bibnamefont
  {Bienfang}}, \bibinfo {author} {\bibfnamefont {A.~J.}\ \bibnamefont {Gross}},
  \bibinfo {author} {\bibfnamefont {A.}~\bibnamefont {Mink}}, \bibinfo {author}
  {\bibfnamefont {B.~J.}\ \bibnamefont {Hershman}}, \bibinfo {author}
  {\bibfnamefont {A.}~\bibnamefont {Nakassis}}, \bibinfo {author}
  {\bibfnamefont {X.}~\bibnamefont {Tang}}, \bibinfo {author} {\bibfnamefont
  {R.}~\bibnamefont {Lu}}, \bibinfo {author} {\bibfnamefont {D.~H.}\
  \bibnamefont {Su}}, \bibinfo {author} {\bibfnamefont {C.~W.}\ \bibnamefont
  {Clark}}, \bibinfo {author} {\bibfnamefont {C.~J.}\ \bibnamefont {Williams}},
  \bibinfo {author} {\bibfnamefont {E.~W.}\ \bibnamefont {Hagley}}, \ and\
  \bibinfo {author} {\bibfnamefont {J.}~\bibnamefont {Wen}},\ }\href {\doibase
  10.1364/OPEX.12.002011} {\bibfield  {journal} {\bibinfo  {journal} {Opt.
  Express}\ }\textbf {\bibinfo {volume} {12}},\ \bibinfo {pages} {2011}
  (\bibinfo {year} {2004})}\BibitemShut {NoStop}%
\bibitem [{\citenamefont {Patel}\ \emph {et~al.}(2012)\citenamefont {Patel},
  \citenamefont {Dynes}, \citenamefont {Choi}, \citenamefont {Sharpe},
  \citenamefont {Dixon}, \citenamefont {Yuan}, \citenamefont {Penty},\ and\
  \citenamefont {Shields}}]{Patel2012}%
  \BibitemOpen
  \bibfield  {author} {\bibinfo {author} {\bibfnamefont {K.~A.}\ \bibnamefont
  {Patel}}, \bibinfo {author} {\bibfnamefont {J.~F.}\ \bibnamefont {Dynes}},
  \bibinfo {author} {\bibfnamefont {I.}~\bibnamefont {Choi}}, \bibinfo {author}
  {\bibfnamefont {A.~W.}\ \bibnamefont {Sharpe}}, \bibinfo {author}
  {\bibfnamefont {A.~R.}\ \bibnamefont {Dixon}}, \bibinfo {author}
  {\bibfnamefont {Z.~L.}\ \bibnamefont {Yuan}}, \bibinfo {author}
  {\bibfnamefont {R.~V.}\ \bibnamefont {Penty}}, \ and\ \bibinfo {author}
  {\bibfnamefont {A.~J.}\ \bibnamefont {Shields}},\ }\href {\doibase
  10.1103/PhysRevX.2.041010} {\bibfield  {journal} {\bibinfo  {journal} {Phys.
  Rev. X}\ }\textbf {\bibinfo {volume} {2}},\ \bibinfo {pages} {041010}
  (\bibinfo {year} {2012})}\BibitemShut {NoStop}%
\bibitem [{\citenamefont {Wang}\ \emph {et~al.}(2017)\citenamefont {Wang},
  \citenamefont {Yin}, \citenamefont {Chau}, \citenamefont {Chen},
  \citenamefont {Wang}, \citenamefont {Guo},\ and\ \citenamefont
  {Han}}]{Wang2017}%
  \BibitemOpen
  \bibfield  {author} {\bibinfo {author} {\bibfnamefont {S.}~\bibnamefont
  {Wang}}, \bibinfo {author} {\bibfnamefont {Z.-Q.}\ \bibnamefont {Yin}},
  \bibinfo {author} {\bibfnamefont {H.}~\bibnamefont {Chau}}, \bibinfo {author}
  {\bibfnamefont {W.}~\bibnamefont {Chen}}, \bibinfo {author} {\bibfnamefont
  {C.}~\bibnamefont {Wang}}, \bibinfo {author} {\bibfnamefont {G.-C.}\
  \bibnamefont {Guo}}, \ and\ \bibinfo {author} {\bibfnamefont {Z.-F.}\
  \bibnamefont {Han}},\ }\href@noop {} {\bibfield  {journal} {\bibinfo
  {journal} {arXiv:1707.00387}\ } (\bibinfo {year} {2017})}\BibitemShut
  {NoStop}%
\end{thebibliography}

%%%%%%%%%%%%%%%%%%%%%%%%%%%% SUPPLEMENT
\onecolumngrid
\clearpage

\begin{center}
\textbf{\large Reconfigurable, single-mode mutually unbiased bases for time-bin qudits: Supplementary material}
\end{center}

\setcounter{equation}{0}
\setcounter{figure}{0}
\setcounter{table}{0}
\makeatletter
\renewcommand{\theequation}{S\arabic{equation}}
\renewcommand{\thefigure}{S\arabic{figure}}

\section{Optimal Solutions}
\label{Sec1}
As noted in the main text, the goal of our simulations is to determine, for a given dimension $d$, time-bin transformations $W^{(m)}$ ($m=0,1,...,d$) that are mutually unbiased within the $d$-mode subspace of interest. After truncating the in general infinite-dimensional space to $S$ modes ($S\gg d$), and considering $N$ unit cells (cf. Fig.~1 of the main text), we can express each transformation $W^{(m)}$ as the matrix product
\begin{equation}
\label{eS1}
W^{(m)} = C_N D_N^{(m)} \cdots C_1 D_1^{(m)}.
\end{equation}
Here, each matrix $C_n$ signifies a FBG with a particular complex reflection pattern, while $D_n^{(m)}$ is a diagonal unitary matrix denoting an EOM that imparts an arbitrary phase shift to each temporal mode.

In contrast to the FBG matrices, the EOM matrices are delineated by basis $m$. This follows from the rapid reconfigurability of the modulation patterns applied to each EOM, which can be updated quickly to switch the measurement basis. On the other hand, the FBGs are constrained to fixed responses across bases. Each matrix $D_n^{(m)}$ is thus parameterized by a real phase function $\varphi_n^{(m)}[l]$ such that
\begin{equation}
\label{eS2}
\left( D_n^{(m)} \right)_{lk} = e^{i \varphi_n^{(m)}[l]} \delta_{lk},
\end{equation}
with $\delta_{lk}$ the Kronecker delta function, nonzero only when $l=k$.

Although the FBG matrices $C_n$ can be decomposed into complex reflections directly---\textit{i.e.}, as $\left( C_n \right)_{lk}=(c_n)_{l-k}$---it is simpler to parameterize $C_n$ in the frequency domain through the transformation
\begin{equation}
\label{eS3}
C_n = F^\dagger \tilde{D}_n F,
\end{equation}
where $F$ is the $S$-point discrete Fourier transform, and
\begin{equation}
\label{eS4}
\left( \tilde{D}_n \right)_{lk} = e^{i \phi_n[l]} \delta_{lk}.
\end{equation}
is a diagonal matrix of spectral phases.  The corresponding complex reflection amplitudes (chips) are then related to the spectral phases via
\begin{equation}
\label{eS5}
c_n [l] = \frac{1}{S} \sum_{s=0}^{S-1} e^{i \phi_n[s]} e^{2\pi i s l/S},
\end{equation}
so that $\sum_s c_n^*[l-s] c_n[k-s] = \delta_{lk}$ (unitarity) follows automatically.

In this formulation, each FBG and EOM configuration is characterized by $S$ real numbers, for a total of $SN(d+2)$ free parameters, after accounting for the $d+1$ basis choices. The optimality of a given set of matrices $W^{(m)}$ is determined by computing the ``mubness'' mean-squared-error $\epsilon_\mathrm{MSE}$ [cf. Eqs. (2)-(4) of the main text].

\begin{figure*}[b]
\includegraphics[width=6.5in]{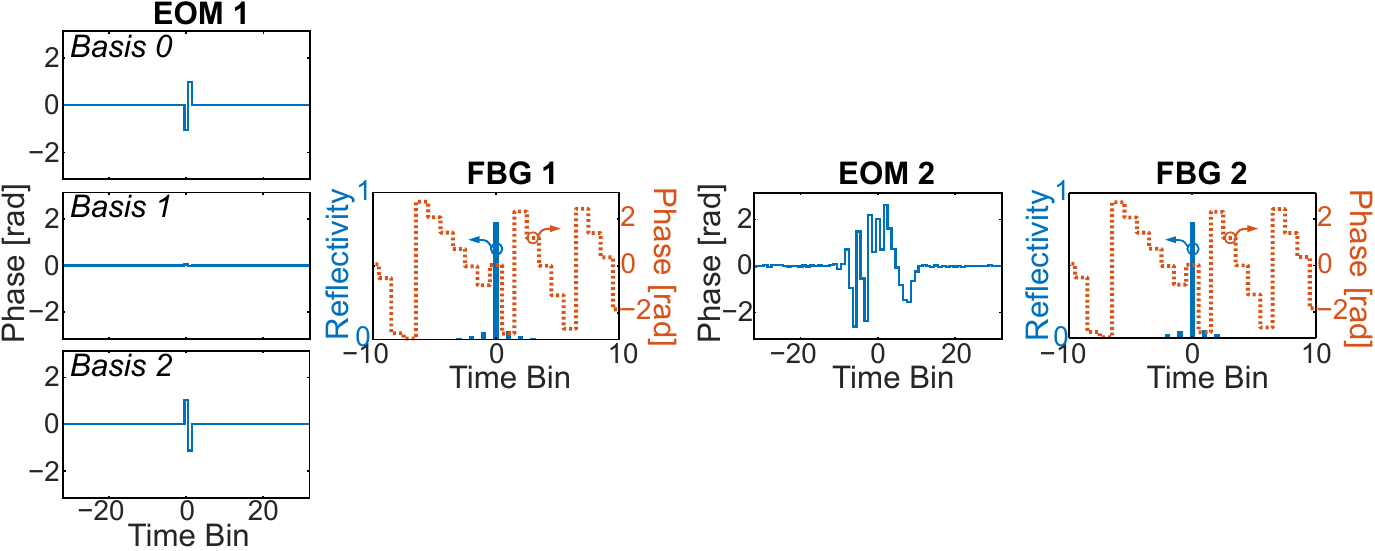}
\caption{Phase modulation patterns and impulse responses for each EOM and FBG, respectively, for measuring three MUBs for $d=2$ time bins. The modulation applied to EOM 1 changes from basis to basis, whereas all other transformations are fixed.}
\label{figS1}
\end{figure*}

\begin{figure*}[!b]
\includegraphics[width=6.5in]{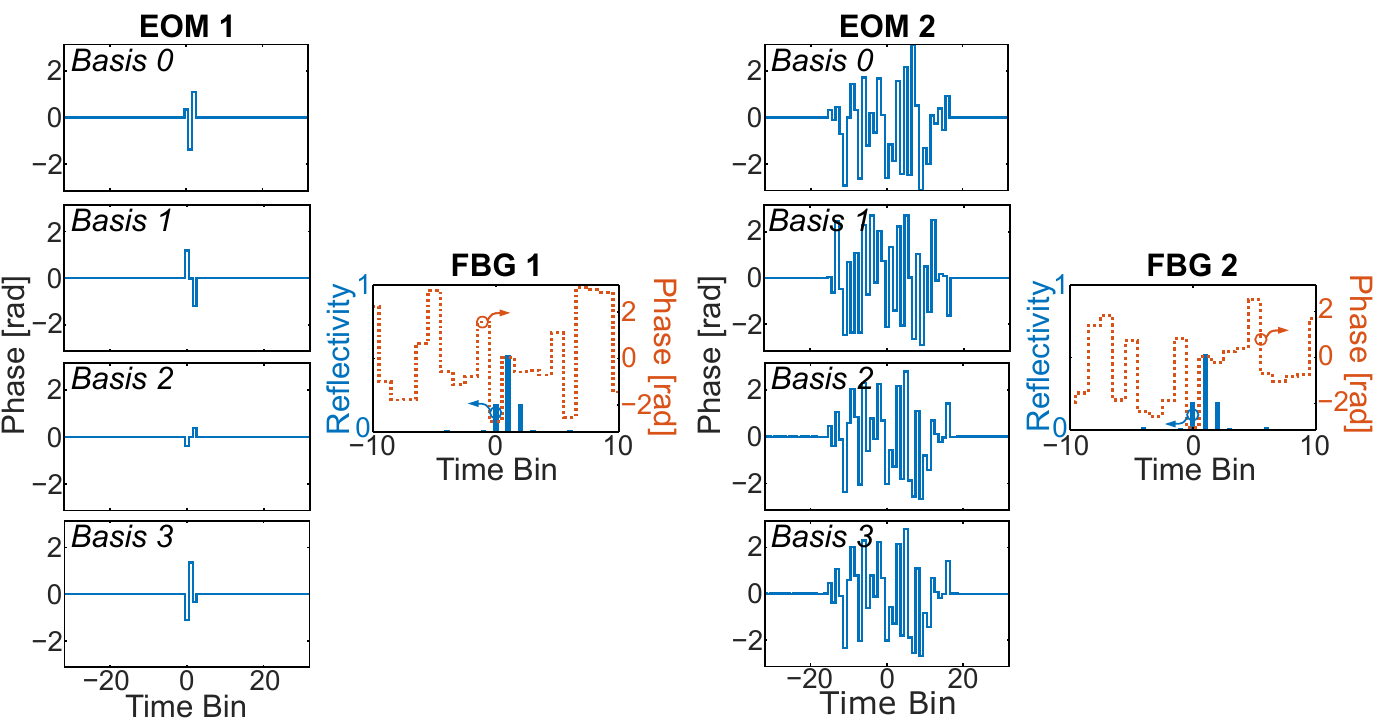}
\caption{Modulation functions and impulse responses to realize four MUBs on time-bin qutrits ($d=3$). The phase modulation patterns applied to both EOM 1 and EOM 2 are selected to produce a particular basis measurement.}
\label{figS2}
\end{figure*}

Before delving into the specific solution forms, we note that embedded within this general formulation are several important physical assumptions. First, the electro-optic phase modulation must be fast enough to apply completely arbitrary phases to successive bins.  Similarly, the pulse duty cycle must be small enough to avoid any distortion effects from the finite electro-optic phase transition slope. Combined, these conditions ensure that fully arbitrary interpulse phases are realizable, while preserving intrapulse uniformity.

Second, the FBG reflection bandwidth must be wide enough to encompass that of each pulse (otherwise additional loss will result). With this assumption fulfilled, the impulse response delays and phase shifts each pulse without temporal distortion.

Finally, the truncation of the full time-bin Hilbert space to $S$ modes must be large enough to approximate infinity, thereby avoiding numerically-induced artifacts resulting from undersampling. In practical terms, any legitimate solution must not contain coupling from the $d$ computational input modes to the exterior modes of the clipped space, which we can confirm \emph{a posteriori} via numerical filtering (see Sec.~\ref{Sec3} below).

For our solutions, we take $S=128$, set the number of EOM/FBG unit cells to $N=2$, and numerically search over all possible phases to minimize $\epsilon_\mathrm{MSE}$. The solution obtained for $d=2$ is shown in Fig.~\ref{figS1}, which reaches $\epsilon_\mathrm{MSE}=1.05\times 10^{-7}$. The temporal phase shifts for each EOM and basis ($\varphi_n^{(m)}[l]$) are shown directly.  For the FBGs, we highlight the discrete-time impulse responses $c_n [l]$, both reflectivity $|c_n [l]|^2$ and phase $\arg c_n [l]$. Interestingly, although we permit three distinct phase patterns for EOM 2 (one for each basis), the solver converges to a configuration with identical modulation across basis choices. Thus, only EOM 1 must be updated to shift between MUBs, making the solution even simpler than demanded physically. Additionally, we note that the FBG impulse responses are not causal, with appreciable reflectivities at negative delays; this simply means that an overall delay must be present in practice, the amount equal to roughly half of the total number of chips kept in the implementation (quantified in Sec.~\ref{Sec3}).

\begin{figure*}
\includegraphics[width=6.5in]{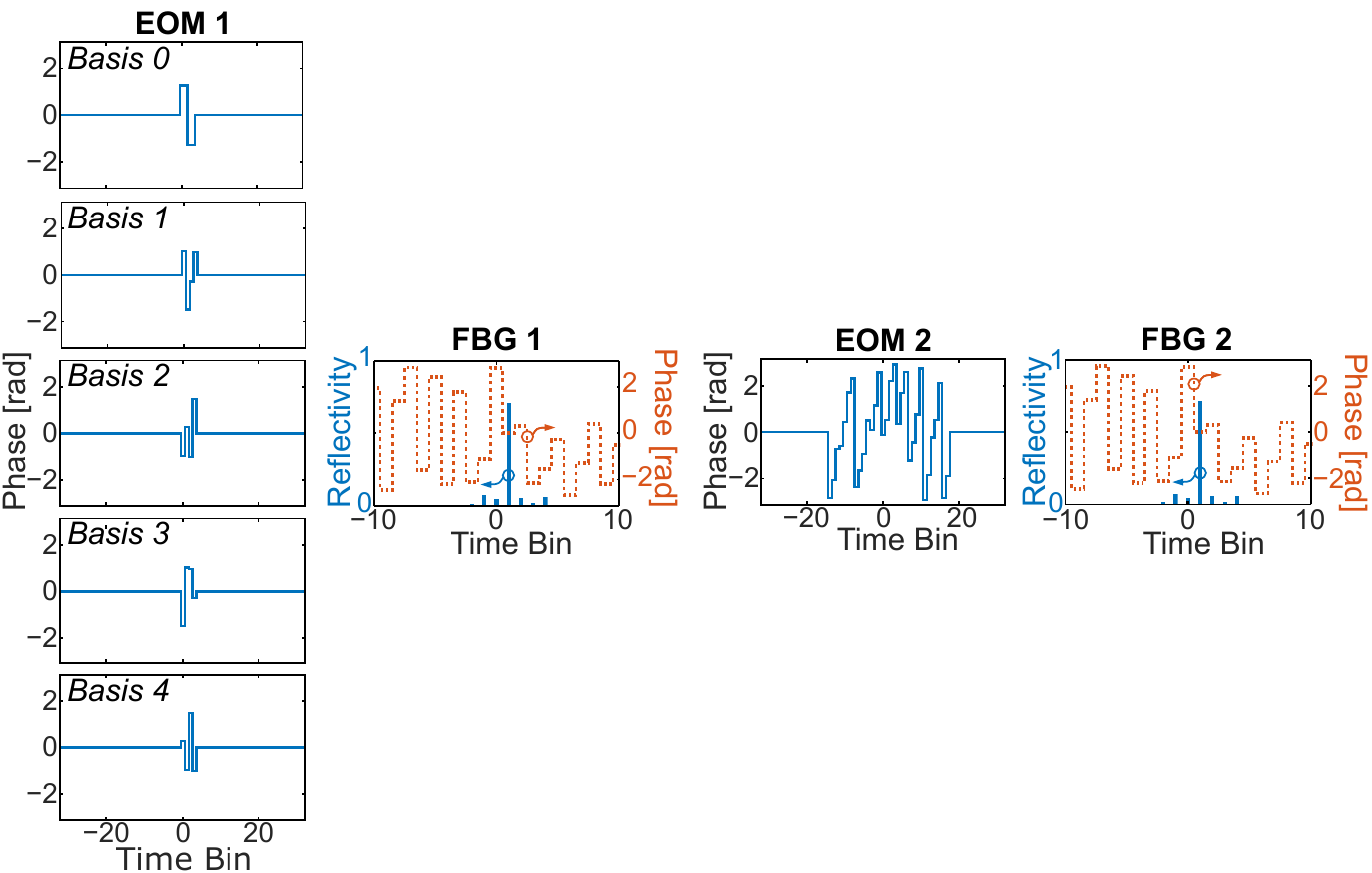}
\caption{EOM and FBG transformations producing MUBs for $d=4$. By selecting the particular phase modulation pattern for EOM 1, any of five mutually unbiased measurements can be effected.}
\label{figS3}
\end{figure*}

The solution for $d=3$ follows in Fig.~\ref{figS2} ($\epsilon_\mathrm{MSE}=1.50\times 10^{-4}$). One more MUB is available, and unlike the $d=2$ case, EOM 2's phases must be modified from basis-to-basis as well as those of EOM 1 (no simplification similar to that of $d=2$ was found). Rounding out the optimizations, Fig.~\ref{figS3} furnishes the modulation functions for the five MUBs in $d=4$, which together enable $\epsilon_\mathrm{MSE}=1.60 \times 10^{-4}$. As with the $d=2$ case, the second EOM is fixed across bases, demanding only that EOM 1 be modified from measurement to measurement.

\section{Basis States}
\label{Sec2}
We next examine the basis states corresponding to the mode transformations described above. As specified by Eq.~(5) of the main text, we define the normalized state vectors $|\nu_m [n]\rangle$ ($m=0,1,...,d$; $n=0,1,...,d-1$) such that applications of transformation $W^{(m)}$ ideally projects $|\nu_m [n]\rangle$ onto the pure time bin $|n\rangle$, whereas $W^{(m^\prime)}$ ($m^\prime \neq m$) produces an equi-amplitude $d$-mode time-bin superposition. In this way, all detections occur within the time-bin eigenbasis, even though the input states themselves may possess a nontrivial phase/amplitude structure.

\begin{figure}
\includegraphics[width=3.4in]{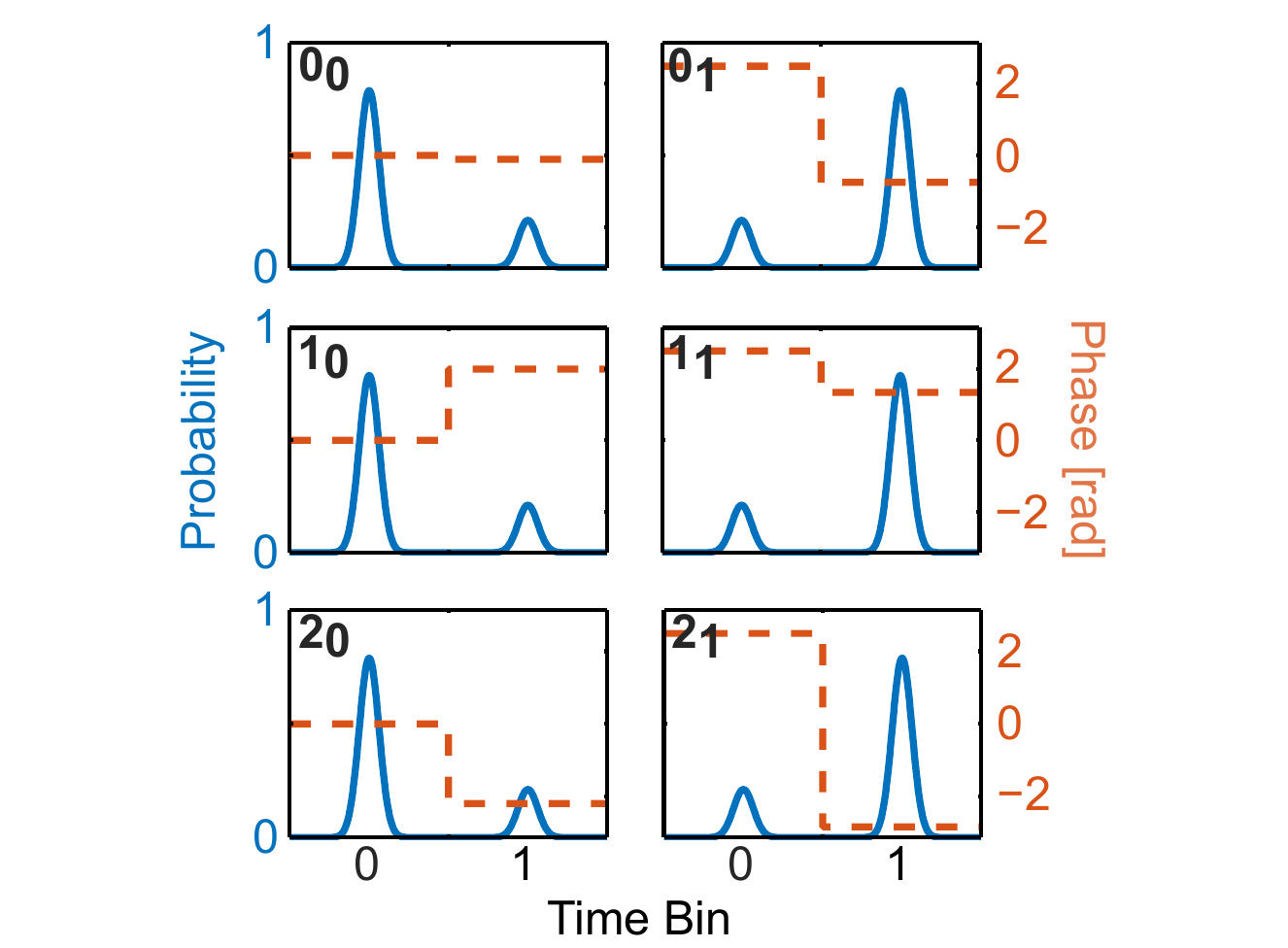}
\caption{Basis states for $d=2$. Each row denotes one of the three MUBs; subscripts label the particular states.}
\label{figS4}
\end{figure}

For example, Fig.~\ref{figS4} shows the six basis states for $d=2$. The probability and phase are shown as functions of time bin, with a finite pulse width provided for aesthetics. The label pair $m_n$ marks state $n$ of basis $m$ ($|\nu_m [n]\rangle$). Because we enforce no \emph{a priori} requirement that one MUB be the time-bin basis (the optimizer is free to choose), all states show finite probability in both basis states; this poses no serious technical limitation in state preparation, since amplitude and phase can be readily set by modulators. The greater challenge is the basis measurement---precisely what this EOM/FBG setup accomplishes.

\begin{figure}
\includegraphics[width=3.4in]{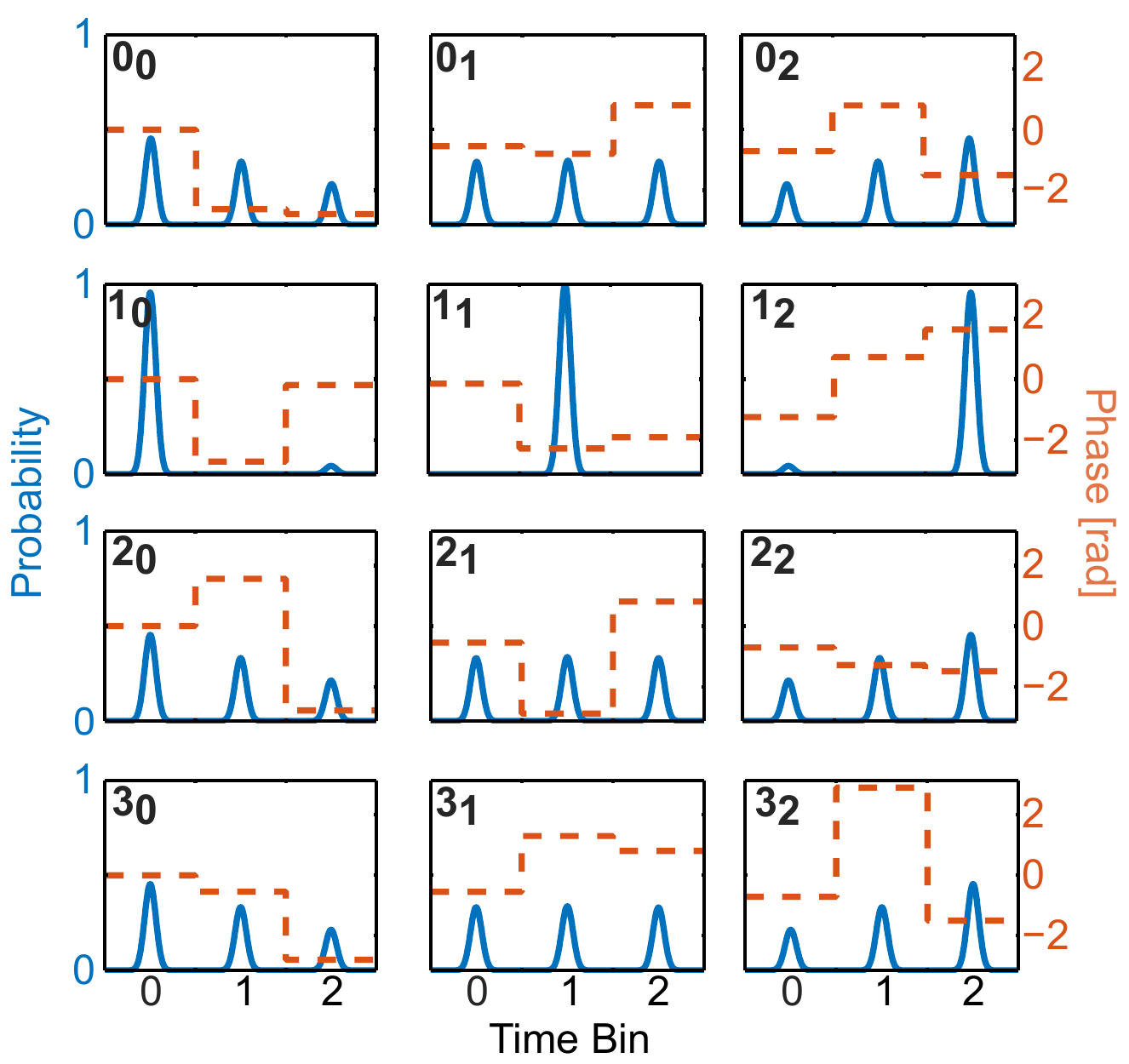}
\caption{Basis sates for $d=3$.}
\label{figS5}
\end{figure}

\begin{figure*}
\includegraphics[width=6.5in]{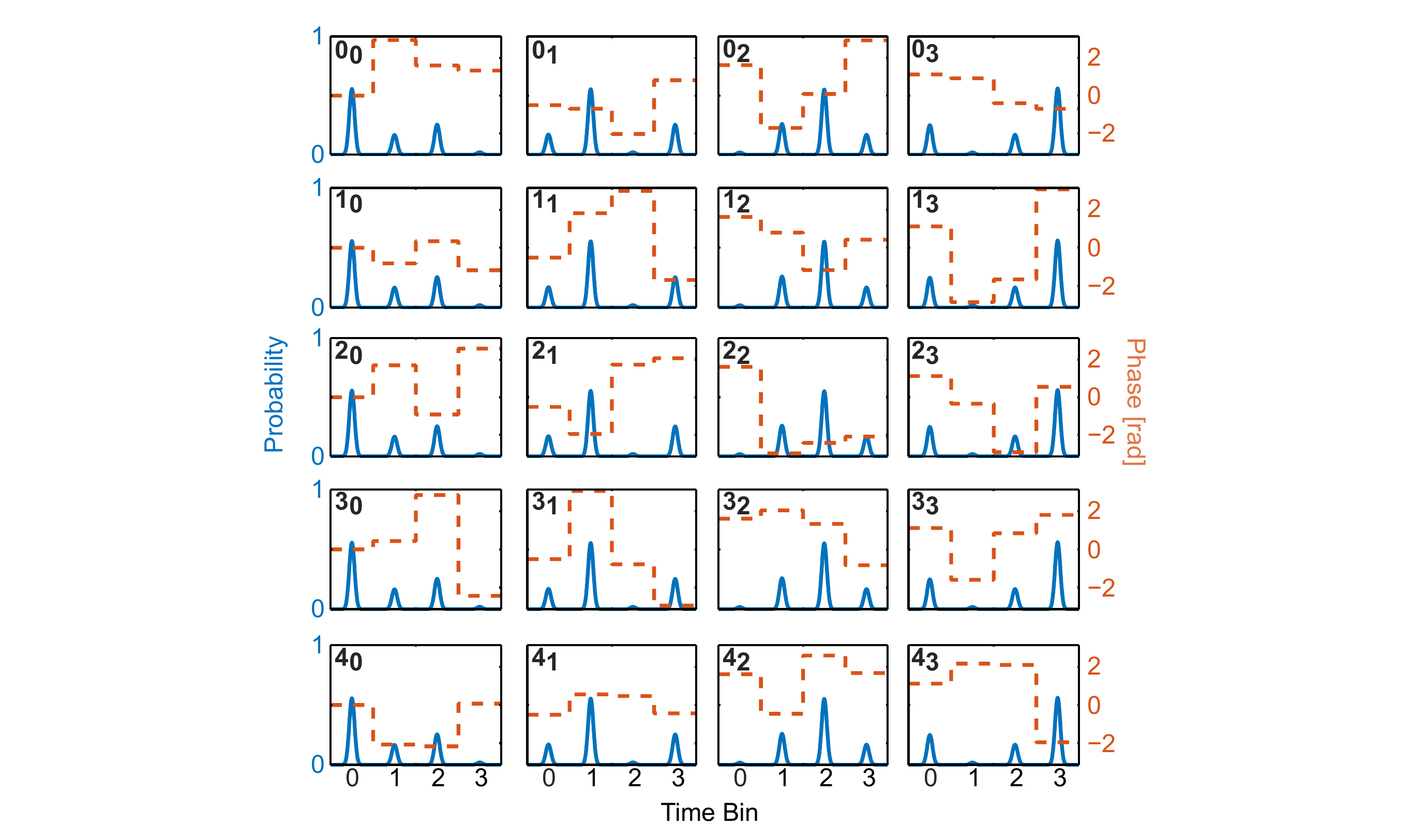}
\caption{Basis states for $d=4$.}
\label{figS6}
\end{figure*}

Figures \ref{figS5} and \ref{figS6} show the basis states for the $d=3$ and $d=4$ solutions, respectively. The same considerations and formalism for $d=2$ apply here as well.

\section{Detection Probabilities}
As noted in the main text, single-photon states prepared according to each solution (from Figs.~\ref{figS4}-\ref{figS6}) and sent through the relevant EOM/FBG network (Figs.~\ref{figS1}-\ref{figS3}) are projected onto time-bin states that signify the $d$ possible outcomes in the chosen measurement basis. For a given state selection $|\nu_m[n]\rangle$ and measurement $V^{(p)}$, we quantify performance via two metrics: (i) the detection probability $\mathcal{D}_{pm}[n]$, which gives the probability the photon indeed remains in the $d$-dimensional output subspace; and (ii) the post-selected probability $\mathcal{P}_{pm}[q|n]$ that, given the photon does remain in the desired subspace, it is found in time-bin $q$.

\begin{figure*}
\includegraphics[width=6.5in]{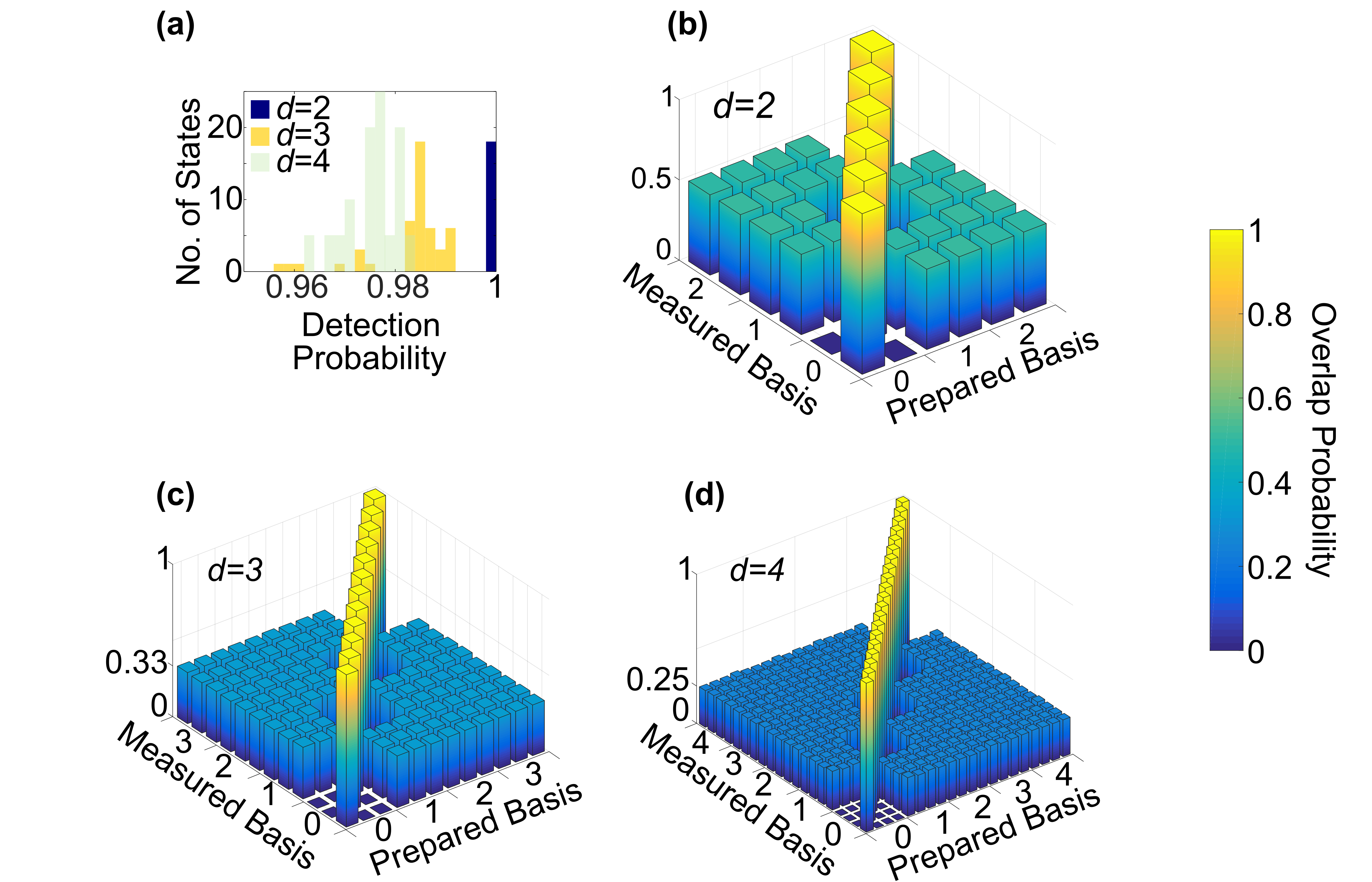}
\caption{(a) Histogram of detection probabilities $\mathcal{D}_{pm}[n]$ for all state/measurement basis combinations. Post-selected probability distributions $\mathcal{P}_{pm}[q|n]$ for (b) $d = 2$, (c) $d = 3$, and (d) $d = 4$. The tick marks at $1/d$ show the expected level for mismatched prepare/measure basis choices.}
\label{figS65}
\end{figure*}

Figure~\ref{figS65} shows these values for all three solutions ($d=2,3,4$). Panels (a) and (b) are reproduced from Fig.~3 of the main text, but are included here to aid in comparison of the results for all dimensions. The histogram of Fig.~\ref{figS65}(a) bins the $d(d+1)^2$ detection probabilities [$d(d+1)$ total basis states $\times$ $(d+1)$ measurement choices]. The case $d=2$ is closest to the ideal, with $d=3,4$ still attaining values of $\mathcal{D}_{pm}[n]$ above $\sim$0.96. The post-selected probabilities for $d=2,3,4$ follow in Figs.~\ref{figS65}(b)-(d). For all three cases, matched basis choices produce localized probability peaks, as expected for near-perfect correlation. Mismatched choices leave uniform outcomes for each input state, all with probabilities around $1/d$. These probability sets form the foundation of our estimates of QKD performance.

\section{Chip Number Analysis}
\label{Sec3}
In evaluating the aforementioned solutions in the context of practical implementation, an important consideration is the effective number of chips needed on each FBG to realize full functionality. For while parameterizing the FBG transformation in the frequency domain [Eq.~(\ref{eS3})] proves extremely convenient for enforcing unitarity, the resulting impulse response  can potentially extend over the entire numerical domain. To monitor against such a situation, here we intentionally remove chips from the FBG responses and quantify the impact on basis measurement performance. Such tests serve a twofold purpose: (i) they validate the solution's trustworthiness against potential aliasing effects resulting from numerical truncation; (ii) they reveal precisely just how many chips are required from fabricated FBGs.

\begin{figure}
\includegraphics[width=4.in]{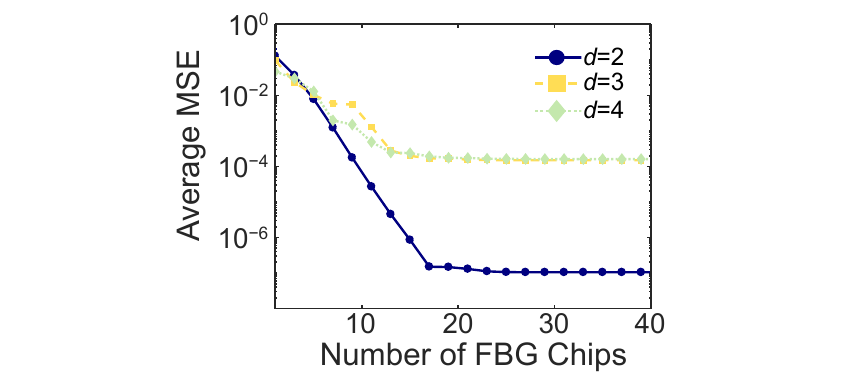}
\caption{Average mean-squared error ($\epsilon_\mathrm{MSE}$) as a function of the number of chips applied by each coded FBG.}
\label{figS7}
\end{figure}

For our particular solutions, we force all but some subset of FBG chips to zero and recompute the average MSE ($\epsilon_\mathrm{MSE}$). Starting with only one nonzero chip (time-bin 0 in the FBG plots above), we successively increase the number of chips in increments of two---adding one preceding time bin and one succeeding. The results of these tests for all dimensions $d=2,3,4$ are given in Fig.~\ref{figS7}. As expected, the error begins high and drops off rapidly, until leveling off at the respective optima. The error floors are reached well before incorporating the full numerical size (128 chips), confirming the validity of the solutions in light of modal truncation. Finally, for the purposes of design, we observe that including any chips beyond $\sim$20 is superfluous in all cases, thus providing a quantitative bound on required complexity. We make use of this number in the main text to estimate FBG length in view of the expected time-bin spacing.

\end{document}